\documentstyle[12pt]{article}

\begin{document}

\vspace{2mm}

\begin{flushright}
MRI-PHY/96-35 \\
hep-th/9611223
\end{flushright}

\vspace{2ex}

\begin{center}
{\large \bf 
Early Universe Evolution in Graviton-Dilaton Models}

\vspace{6mm}
{\large  S. Kalyana Rama}
\vspace{3mm}

Mehta Research Institute, 10 Kasturba Gandhi Marg, 

Allahabad 211 002, India. 

\vspace{1ex}
email: krama@mri.ernet.in \\ 
\end{center}

\vspace{4mm}

\begin{quote}
ABSTRACT. 
We present a class of graviton-dilaton models which leads 
to a singularity free evolution of the universe. We study 
the evolution of a homogeneous isotropic universe. We 
follow an approach which enables us to analyse 
the evolution and obtain its generic features even in 
the absence of explicit solutions, which are not possible 
in general. We describe the generic evolution of 
the universe and show, in particular, that it is 
singularity free in the present class of models. 
Such models may stand on their own as interesting 
models for singularity free cosmology, and may be 
studied accordingly. They may also arise from string 
theory. We discuss critically a few such possibilities.  
\end{quote}

\newpage

\begin{center}
{\bf 1. Introduction} 
\end {center} 


General Relativity (GR) is a beautiful theory and, more 
importantly, has been consistently successful in describing 
the observed universe. But, GR leads inevitably to 
singularities. In particular, our universe according to GR 
starts from a big bang singularity. We do not, however, 
understand the physics near the singularity nor know its 
resolution. The inevitability and the lack of understanding 
of the singularities point to a lacuna in our fundamental 
understanding of gravity itself. By the same token, 
a successful resolution of the singularities 
is likely to provide deep insights into gravity. 

In order to resolve the singularities, it is necessary 
to go beyond GR, perhaps to a quantum theory of gravity. 
A leading candidate for such a theory is string theory. 
However, despite enormous progress \cite{rev}, string 
theory has not yet resolved the singularities \cite{vafa}. 

Perhaps, one may have to await further progress in 
string theory, or a quantum theory of gravity, to be able 
to resolve singularities. However, there is still an avenue 
left open - resolve the singularities, if possible, within 
the context of a generalised Brans-Dicke model, and then 
inquire whether such a model can arise from string theory, 
or a quantum theory of gravity. 

Generalised Brans-Dicke model is special for more than 
one reason. It appears naturally in supergravity, 
Kaluza-Klein theories, and in all the known effective 
string actions. It is perhaps the most natural extension 
of GR \cite{will}, which may explain its ubiquitous 
appearance in fundamental theories: Dicke had discovered 
long ago the lack of observational evidence for 
the principle of strong equivalence in GR. Together 
with Brans, he then constructed the Brans-Dicke theory 
\cite{bd} which obeys all the principles of GR except 
that of strong equivalence. This theory, generalised 
in \cite{gbd}, has been applied in many cosmological 
and astrophysical contexts \cite{mb}-\cite{barrow}. 

Brans-Dicke theory contains a graviton, a scalar (dilaton) 
coupled non minimally to the graviton, and a constant 
parameter $\omega$. GR is obtained when $\omega = \infty$. 
In the generalised Brans-Dicke theory, referred to as 
graviton-dilaton theory, $\omega$ is an arbitrary function 
of the dilaton \cite{will,gbd}. Hence, it includes 
an infinite number of models, one for every function 
$\omega$. Thus, very likely, it also includes all 
effective actions that may arise from string theory 
or a quantum theory of gravity. 

This perspective logically suggests an avenue in resolving 
the singularities: to find, if possible, the class of 
graviton-dilaton models in which the evolution of 
universe is singularity free, and then study whether 
such models can arise from string theory or a quantum 
theory of gravity. On the other hand, even if not 
deriveable from string theory or a quantum theory of 
gravity, such models may stand on their own 
as interesting models for singularity free cosmology. 
In either case, one may study further implications 
of these models in other cosmological and astrophysical 
contexts. Such studies are fruitful and are likely to 
lead to novel phenomena, providing valuable insights. 

Accordingly, in this paper, we consider a class of 
graviton-dilaton models where the function $\omega$ 
satisfies certain constraints. These constraints were 
originally derived in \cite{k1} in a different context, 
and some of their generic cosmological and astrophysical 
consequences were explored in \cite{k1,k2}. Here we 
analyse the evolution of universe in these models which, 
as we will show, turns out to be singularity free. 
The essential points of the analysis and the results 
have been given in a previous letter \cite{k3}. 

We consider a homogeneous isotropic universe, such as our 
observed one, containing matter. The relevent equations 
of motion cannot, in general, be solved explicitly except 
in special cases \cite{barrow,k2}. Hence, a different 
approach is needed for the general case which is valid 
for any matter and for any arbitrary function $\omega$, 
and which enables one to analyse the evolution and obtain 
its generic features even in the absence of explicit 
solutions. 

We will present such an approach below. We first present 
a general analysis of the evolution and then apply it 
to describe in detail the evolution of universe in 
the present model. We show, in particular, that 
the constraints on $\omega$ ensure that the evolution 
is singularity free. The universe evolves with no big 
bang or any other singularity and the time continues 
indefinitely into the past and the future. 

An important question to ask, from our perspective, 
is whether a function $\omega$ as required in the present 
model, can arise from a fundamental theory. We discuss 
critically a few such possibilities in string theory. 
On the other hand, the present model may stand on its 
own as an interesting model for singularity free 
cosmology and may be studied accordingly. We mention 
some issues for future study. 

This paper is organised as follows. In section 2, we 
present our model. In section 3, we write its equations 
of motion in various forms, so as to facilitate our 
analysis. In section 4, we present 
the general analysis of the evolution. In section 5, we 
illustrate our method by applying the results of section 
4 to describe the evolution of toy universes and show 
that their evolution is singularity free in the present 
model. In section 6, we describe the evolution of our 
observed universe and show that its evolution is 
singularity free in the present model. In section 7, we 
discuss further generalisations of our model. In section 
8, we give a brief summary, discuss critically a few 
possibilities of our model arising from string theory, 
and mention some issues for future study. In 
the Appendix, we derive the sufficiency conditions for 
the absence of singularities which are used in the paper. 


\begin{center}
{\bf 2. Graviton-Dilaton Model} 
\end {center} 


We consider the following graviton-dilaton action 
in `Einstein frame': \footnote{In our notation, 
the signature of the metric is 
$(- + + +)$ and $\bar{R}_{\mu \nu \lambda \tau} = 
\frac{1}{2} \frac{\partial^2 \bar{g}_{\nu \lambda}} 
{\partial x^{\mu} \partial x^{\tau}} + \cdots$.} 
\begin{equation}\label{sein} 
S = \frac{1}{16 \pi G_N} \int d^4 x \sqrt{- \bar{g}} 
\left( - \bar{R} + \frac{1}{2} (\bar{\nabla} \phi)^2 
\right) + S_M ({\cal M}, e^{- \psi (\phi)} 
\bar{g}_{\mu \nu}) \; , 
\end{equation}
where $G_N$ is the Newton's constant, 
$\phi$ is the dilaton and $\psi (\phi)$ is 
an arbitrary function that charcterises the theory. $S_M$ 
is the action for matter fields, denoted collectively by 
${\cal M}$. They couple to the metric $\bar{g}_{\mu \nu}$ 
minimally and to the dilaton $\phi$ through the function 
$\psi (\phi)$. This function cannot be gotten rid of 
by a redefinition of $\bar{g}_{\mu \nu}$ except when 
the matter action $S_M$ is conformally invariant - in 
that case, by definition, 
$S_M ({\cal M}, e^{- \psi (\phi)} \bar{g}_{\mu \nu}) 
= S_M ({\cal M}, \bar{g}_{\mu \nu})$ - which is assumed 
to be not the case here. 

We can define a metric 
\begin{equation}\label{conf}
g_{\mu \nu} = e^{- \psi} \bar{g}_{\mu \nu} \; , 
\end{equation}
and write the action (\ref{sein}) in `Dicke frame': 
\begin{equation}\label{spsi} 
S = \frac{1}{16 \pi G_N} 
\int d^4 x \sqrt{- g} e^{\psi} \left( - R 
+ \frac{1}{2} \left(3 \psi_{\phi}^2 - 1 \right)  
(\nabla \phi)^2 \right) + S_M ({\cal M}, g_{\mu \nu})  \; , 
\end{equation}
where $\psi_{\phi} \equiv \frac{d \psi}{d \phi}$. If 
$\psi$ has a finite upper bound $\psi \le \psi_{max} 
< \infty$, then the factor $e^{\psi_{max}}$ can be 
absorbed into $G_N$ and the range of $\psi$ can be set 
to be $\psi \le 0$. Also, we will work in the units where 
$G_N = \frac{e^{\psi_{max}}}{8 \pi}$. 

Both forms of the action in (\ref{sein}) and (\ref{spsi}) 
are, however, equivalent \cite{ein}. 
In `Einstein frame', the matter 
fields feel, besides the gravitational force, 
the dilatonic force also which must be taken into account 
in obtaining the physical quantities. Whereas, in 
`Dicke frame', the matter fields feel only 
the gravitational force and, hence, the physical 
quantities are directly obtained from the metric 
$g_{\mu \nu}$. On the other hand, equations of motion 
are often easier to solve if $\bar{g}_{\mu \nu}$ is used. 

Defining a new dilaton field $\chi = e^{\psi}$, 
the action (\ref{spsi}) can be written as 
\begin{equation}\label{schi}
S = \frac{1}{16 \pi G_N} 
\int d^4 x \sqrt{- g} \left( - \chi R 
+ \frac{\omega (\chi)}{\chi} \; (\nabla \chi)^2 \right) 
+ S_M ({\cal M}, \; g_{\mu \nu})  \; , 
\end{equation}
where $\omega (\chi)$ is now the arbitrary function 
that characterises the theory. This is the form of 
the action used more commonly and, hence, we will also 
use it in this paper in most of what follows. Note that 
$\chi \ge 0$ since $\chi = e^{\psi}$. Furthermore, if 
$\chi$ has a finite upper bound 
$\chi \le \chi_{max} < \infty$, then the factor 
$\chi_{max}$ can be absorbed into $G_N$ and the range of 
$\chi$ can be set to be $0 \le \chi \le 1$. Also, we will 
work in the units where $G_N = \frac{\chi_{max}}{8 \pi}$. 

Let $\Omega \equiv 2 \omega + 3$. Then, $\phi$ 
and $\psi (\phi)$ are related to $\chi$ and 
$\Omega (\chi)$ as follows: 
\begin{equation}\label{omega}
\Omega = \psi_{\phi}^{- 2} 
\; \; \; \;  {\rm or, \; equivalently,} \; \; \; \; 
\phi = \int \frac{d \chi}{\chi} \; 
\sqrt{\Omega (\chi)} \; . 
\end{equation}
Note that $\Omega$ must be positive. 
If one starts from $\phi$ and $\psi (\phi)$, then 
the first relation gives $\Omega$ in terms of $\phi$. 
To obtain $\Omega$ in terms of $\chi, \; \psi (\phi)$ 
must be inverted first to obtain $\phi$ in terms of 
$\psi$ and then $\psi = {\rm ln} \chi$ must be used. 
If one starts from $\chi$ and 
$\Omega (\chi)$ and uses $\chi = e^{\psi}$, then 
the second relation gives $\phi$ in terms of $\psi$ 
which must then be inverted to obtain $\psi$ in terms 
of $\phi$. However, in the following, no inversion is 
necessary since no explicit forms for $\Omega (\chi)$ 
and $\psi (\phi)$ will be used. 

In the model we study here, the function $\psi (\phi)$ 
is required to satisfy the following constraints: 
\begin{eqnarray}
\lim_{|\phi| \to \infty} \psi & = & 
- \frac{|\phi|}{\sqrt{\Omega_0}}, 
\; \; \; \; \Omega_0 \le \frac{1}{3} \nonumber \\ 
\frac{d^n \psi}{d \phi^n} & = & {\rm finite} 
\; \; \; \; \forall \; n \ge 1, \; \; \; \; 
- \infty \le \phi \le \infty \nonumber \\
\frac{d \psi}{d \phi} & = & 0 \; \; \; \; {\rm at} 
\; \; \phi = 0 \; \; {\rm only} \nonumber \\ 
\lim_{\phi \to 0} \psi (\phi) & = & - k \phi^{2 n} 
\; , \; \; \; \; n \ge 1 \; , \label{12phi} 
\end{eqnarray}
where $\Omega_0 > 0$ and $k > 0$ are constants and $n$ 
is an integer. The function $\psi (\phi)$ is otherwise 
arbitrary. 

The first two constraints are important and were 
originally obtained in a different context in 
\cite{k1,k2}, where some of their generic consequences 
were also explored. These constraints imply 
a finite upper bound on $\psi$, {\em i.e.}\ $\psi (\phi) 
\le \psi_{{\rm max}} < \infty$. Hence, the factor 
$e^{\psi_{max}}$ can be absorbed into $G_N$ and the range 
of $\psi$ can be set to be $\psi \le 0$. Thus there is 
atleast one critical point where $\frac{d \psi}{d \phi} 
= 0$, and $\psi = 0$. 

The third constraint is not important, but assumed here 
for the sake of simplicity only. It implies that there 
is only one critical point, a maximum, of $\psi (\phi)$. 
As explained above, we can take $\psi_{max} = 0$. Thus, 
the range of $\psi$ is given by $- \infty \le \psi \le 0$ 
and, hence, that of $\chi = e^{\psi}$ by 
$0 \le \chi \le 1$. Also, by adding a suitable constant 
to $\phi$, we can take the maximum of $\psi (\phi)$ to 
occur at $\phi = 0$ with no loss of generality. 

The fourth constraint follows from the Taylor expansion 
of $\psi (\phi)$ near its maximum at $\phi = 0$. Note, 
as follows from (\ref{omega}), that $\Omega \to \infty$ 
as $\phi \to 0$. 

The constraints in (\ref{12phi}), expressed equivalently 
in terms of $\chi$ and $\Omega (\chi)$, become 
\begin{eqnarray}
\Omega (0) & = & \Omega_0 \le \frac{1}{3}  \nonumber \\ 
\frac{d^n \Omega}{d \chi^n} & = & {\rm finite} 
\; \; \; \; \forall \; n \ge 1, \; \; \; \; 
0 \le \chi < 1  \nonumber \\ 
\Omega & \to & \infty \; \; \; \; {\rm at} 
\; \; \chi = 1 \; \; {\rm only} \nonumber \\ 
\lim_{\chi \to 1} \Omega & = & 
\Omega_1 (1 - \chi)^{- 2 \alpha} \; , \; \; \; \; 
\frac{1}{2} \le \alpha < 1 \; , \label{123} 
\end{eqnarray} 
where $\Omega_0 > 0$ and $\Omega_1 > 0$ are constants, 
and the range of $\alpha$ follows naturally from 
(\ref{omega}) and (\ref{12phi}). We further assume that 
$\Omega (\chi)$ is a strictly increasing function of 
$\chi$. This constraint is not important, but assumed 
here for the sake of simplicity only. The function 
$\Omega (\chi)$ is otherwise arbitrary. 

Note, for later reference, that $\Omega \to \infty$ and 
$\frac{1}{\Omega^3} \frac{d \Omega}{d \chi} = 
- \frac{2 \alpha}{\Omega_1^2} (1 - \chi)^{4 \alpha - 1}$ 
in the limit $\chi \to 1$. Hence, for the range 
of $\alpha$ given in (\ref{123}), which follows 
naturally from (\ref{12phi}), we have that 
$\frac{1}{\Omega^3} \frac{d \Omega}{d \chi} \to 0$ 
in the limit $\chi \to 1$. This ensures, as 
will become clear later, that the present model 
satisfies the observational constraints imposed 
by solar system experiments. 

A vast class of functions $\Omega (\chi)$, equivalently 
$\psi (\phi)$, exists satisfying the above constraints. 
A simple example is $\Omega = \Omega_0 + \Omega_1 
((1 - \chi)^{- 2 \alpha} - 1)$. However, no explicit 
form for $\Omega (\chi)$ will be used in the following. 

\begin{center}
{\bf 3. Equations of Motion} 
\end {center} 


In the following, we take the ``matter'' to be a perfect 
fluid with density $\rho$ and pressure $p$, related by 
$p = \gamma \rho$ where $-1 \le \gamma \le 1$. Note that 
the value of $\gamma$ indicates the nature of ``matter''. 
For example, $\gamma = 0, \; \frac{1}{3}$, or $1$ 
indicates that the ``matter'' is dust, radiation, or 
massless scalar field respectively. Hence, we enclose 
``matter'' within inverted commas where necessary in 
order to remind the reader that ``matter'' does not mean 
dust only. In the following, we study the evolution of 
a flat homogeneous isotropic universe. (The following 
analysis can, however, be extended to curved universes 
also and the main results remain unchanged.) The line 
element is given by 
\begin{equation}\label{ds}
d s^2 = - d t^2 + e^{2 A(t)} (d r^2 + r^2 d S_2^2) \; , 
\end{equation} 
where $e^A$ is the scale factor and $d S_2$ is the line 
element on an unit sphere, and the fields depend on 
the time coordinate $t$ only. The equation of motion 
for $A$, following from (\ref{schi}), is given by 
\begin{equation}\label{ad2}
\dot{A}^2 + \frac{\dot{A} \dot{\chi}}{\chi} 
- \frac{\omega \dot{\chi}^2}{\chi^2} 
= \frac{\rho}{6 \chi} \; , 
\end{equation} 
where upper dots denote $t$-derivatives. Solving 
equation (\ref{ad2}) for $\dot{A}$, we obtain 
\begin{equation}\label{ad}
\dot{A} = - \frac{\dot{\chi}}{2 \chi} + \epsilon \; 
\sqrt{\frac{\rho}{6 \chi} + \frac{\Omega \dot{\chi}^2}
{12 \chi^2}} \; , 
\end{equation} 
where $\epsilon = \pm 1$ and we have used $\Omega = 2 
\omega + 3$. The square roots in (\ref{ad}) and in 
the following are to be taken with a positive sign 
always. The remaining equations of motion, following 
from (\ref{schi}), are given by 
\begin{eqnarray} 
\ddot{\chi} + 3 \dot{A} \dot{\chi} 
+ \frac{\dot{\Omega} \dot{\chi}}{2 \Omega} & = & 
\frac{(1 - 3 \gamma) \rho}{2 \Omega} \label{chidd} \\ 
\rho & = & \rho_0 e^{- 3 (1 + \gamma) A} \; , \label{rho} 
\end{eqnarray} 
where we have used $p = \gamma \rho$ and $\rho_0 \ge 0$ 
is a constant. Equations of motion following from 
(\ref{schi}) also contain the equation for $\ddot{A}$: 
\begin{equation}\label{add}
{\cal A} \equiv 2 \ddot{A} + 3 \dot{A}^2 
+ \frac{2 \dot{A} \dot{\chi}}{\chi} 
+ \frac{\omega \dot{\chi}^2}{2 \chi^2} 
+ \frac{\ddot{\chi}}{\chi} + \frac{p}{2 \chi} = 0 \; . 
\end{equation}
In general, this equation follows from (\ref{ad2}), 
(\ref{chidd}), and (\ref{rho}) also. But there is 
a subtlety pointed out in \cite{lessner}. What one 
obtains from (\ref{ad2}), (\ref{chidd}), and (\ref{rho}) 
is the following: \footnote{Although the derivation of 
(\ref{add'}) is straightforward, an outline of the steps 
involved is perhaps helpful. In equation (\ref{chidd}), 
restore the $p$-dependence by replacing $(1 - 3 \gamma) 
\rho$ by $(\rho - 3 p)$. Now, differentiate (\ref{ad2}) 
and use $\dot{\rho} = -  3 \dot{A} (\rho + p)$. Then 
substitute (\ref{ad2}) for $\frac{\rho}{6 \chi}$, and 
(\ref{chidd}) for $\dot{\omega}$, equivalently 
$\dot{\Omega}$. Again substitute (\ref{ad2}) for 
$\frac{\rho}{6 \chi}$. The result is equation 
(\ref{add'}) \cite{lessner}.} 
\begin{equation}\label{add'}
\left( \dot{A} + \frac{\dot{\chi}}{2 \chi} \right) 
{\cal A} = 0 \; , 
\end{equation}
where the expression for ${\cal A}$ is defined in 
(\ref{add}). Equation (\ref{add}) then follows if and 
only if $\left( \dot{A} + \frac{\dot{\chi}}{2 \chi} 
\right) \ne 0$, {\em i.e.} if and only if $e^A$ is 
not proportional to $\chi^{- \frac{1}{2}}$. This subtlety 
should be kept in mind. However, as checked explicitly, 
the solutions presented in this paper all satisfy 
equations (\ref{ad2}) - (\ref{add}). 

Equations (\ref{ad}) and (\ref{chidd}) can also be 
written in different forms. Equation (\ref{chidd}) 
can be integrated once to obtain 
\begin{eqnarray}
\dot{\chi} (t) & = & \frac{e^{- 3 A}}{\sqrt{\Omega}} \; 
(\sigma (t) + c) \label{chid} \\ 
{\rm where} \; \; \; \;    \; \; \; \;   \; \; 
\sigma (t) & = & \frac{(1 - 3 \gamma) \rho_0}{2} 
\int^t_{t_i} dt \frac{e^{- 3 \gamma A}}{\sqrt{\Omega}} 
\label{sigma} 
\end{eqnarray}
and $c = \dot{\chi} \sqrt{\Omega} e^{3 A} \; 
\vert_{t = t_i}$ is a constant. An equation 
relating $A$ and $\chi$ can now be derived. Divide 
equation (\ref{ad}) by $\frac{\dot{\chi}}{\chi}$ and 
substitute (\ref{chid}) for $\dot{\chi}$. We then have 
\begin{equation}\label{achi}
2 \chi \; \frac{d A}{d \chi} = - 1 
+ \epsilon \; {\rm sign} (\dot{\chi}) \; \sqrt{K} \; , 
\; \; \; \; K \equiv \frac{\Omega}{3} \; 
\left( 1 + \frac{2 \rho_0 \chi e^{3 (1 - \gamma) A}}
{(\sigma (t) + c)^2} \right) \; .  
\end{equation}

All the above equations can be equivalently written 
in terms of $\phi$ and $\psi (\phi)$ also. For example, 
equations (\ref{ad}) and (\ref{chidd}) become 
\begin{eqnarray} 
\dot{A} + \frac{\dot{\psi}}{2} & = & \epsilon \; 
\sqrt{\frac{\rho e^{- \psi}}{6} 
+ \frac{\dot{\phi}^2}{12}} \label{adpsi} \\ 
\ddot{\phi} + 3 \dot{A} \dot{\phi} + \dot{\psi} \dot{\phi} 
& = & \frac{(1 - 3 \gamma) \rho}{2} \; \psi_{\phi} 
\; e^{- \psi} \label{phidd} \; , 
\end{eqnarray} 
while equation (\ref{chid}) and (\ref{sigma}) become 
\begin{eqnarray}
\dot{\phi} (t) & = & e^{- 3 A - \psi} \; (\sigma (t) + c) 
\label{phid} \\ {\rm where} \; \; \; \;    \; \; \; \;   
\; \; \sigma (t) & = & \frac{(1 - 3 \gamma) \rho_0}{2} 
\int^t_{t_i} dt \; \psi_{\phi} e^{- 3 \gamma A} 
\label{sigmapsi} 
\end{eqnarray}
and $c = \dot{\phi} e^{3 A + \psi} \; \vert_{t = t_i}$  
is a constant. 

We now make a few remarks. Firstly, under time reversal 
$t \to - t, \;  \epsilon \to - \epsilon$ in equations 
(\ref{ad}) and (\ref{adpsi}), whereas equations 
(\ref{chidd}) and (\ref{phidd}) are unchanged. Thus, 
evolution for $- \epsilon$ is same as that for 
$\epsilon$, but with the direction of time reversed. 

Secondly, in general, there will appear in the solutions 
positive non zero arbitrary constant factors 
$e^{A_0}, \; \rho_0$, and $\chi_0$ in front of 
$e^A, \; \rho$, and $\chi$ respectively. However, it 
follows from (\ref{ad}) and (\ref{chidd}) that they can 
all be set equal to $1$ with no loss of generality if 
time $t$ is measured in units of 
$\sqrt{\frac{\chi_0 e^{3 (1 + \gamma) A_0}}{\rho_0}}$. 
Therefore, in the solutions below we often assume that 
$t$ is measured in appropriate units and, hence, set 
these constants equal to $1$. Note, however, that 
the model dependent constants associated with 
$\Omega (\chi)$, such as $\Omega_0$ or $\Omega_1$ 
in (\ref{123}), cannot be set equal to $1$. 

Thirdly, if $\rho = 0$ or $\gamma = \frac{1}{3}$ then 
$\sigma (t) = 0$ in (\ref{chid}). If it is also 
possible to perform the integrations given below 
then solutions to equations (\ref{ad}) and (\ref{chid}) 
can, in principle, be obtained in a closed form. To 
see this, define two new variables $\bar{t}$ and 
$\bar{\chi}$ as follows: 
\begin{equation}\label{tanew}
d \bar{t} \equiv \sqrt{\chi} d t \; , \; \; \; \; 
e^{\bar{A}} \equiv \sqrt{\chi} e^A \; . 
\end{equation}
Note that these variables appear naturally in `Einstein 
frame' given in (\ref{sein}) and (\ref{conf}). After some 
algebra, equations (\ref{ad}) and (\ref{chid}) become 
\begin{eqnarray}
\frac{d \bar{A}}{d \bar{t}} & = & 
\frac{\epsilon c e^{- 3 \bar{A}}}{\sqrt{12}} \; 
\sqrt{1 + \frac{2 \rho_0 e^{2 \bar{A}}}{c^2}} 
\label{adbar} \\
\frac{1}{\chi} \frac{d \chi}{d \bar{t}} & = & 
\frac{c e^{- 3 \bar{A}}}{\sqrt{\Omega}} \; , 
\label{chidbar} 
\end{eqnarray} 
from which it follows that 
\begin{equation}\label{achibar}
\frac{\chi d \bar{A}}{d \chi} = \epsilon \; 
\sqrt{\frac{\Omega}{12} \; \left( 
1 + \frac{2 \rho_0 e^{2 \bar{A}}}{c^2} \right)} \; . 
\end{equation} 
In the above equations $\epsilon = \pm 1$ as in 
(\ref{ad}) and $c$ is a constant as in (\ref{chid}). 

It turns out that $\bar{A}$-integrations in equations 
(\ref{adbar}) and (\ref{achibar}) can be performed 
explicitly. Omitting the details, which are 
straightforward, the result is as follows. When 
$\rho$ and, hence, $\rho_0 = 0$: 
\begin{eqnarray}
e^{3 \bar{A}} & = & \frac{\epsilon c \sqrt{3}}{2} 
\int d \bar{t} \label{avst0} \\
\bar{A} & = & \epsilon \int \frac{d \chi}{\chi} \; 
\sqrt{\frac{\Omega}{12}} \; . \label{avschi0} 
\end{eqnarray} 
When $\rho \ne 0$, define $y \equiv 
\sqrt{\frac{2 \rho_0}{c^2}} \; e^{\bar{A}}$. Then 
\begin{eqnarray} 
y \sqrt{1 + y^2} - {\rm ln} (\sqrt{1 + y^2} + y) & = & 
\frac{\epsilon c}{\sqrt{3}} \left( \frac{2 \rho_0}{c^2} 
\right)^{\frac{3}{2}} \; \int d \bar{t} \label{avst} \\
{\rm ln} y - {\rm ln} (\sqrt{1 + y^2} + 1) & = & 
\epsilon \int \frac{d \chi}{\chi} \; 
\sqrt{\frac{\Omega}{12}} \; . \label{avschi} 
\end{eqnarray} 

One thus has an explicit relation between $\bar{A}$ and 
$\bar{t}$, and also between $\bar{A}$ and $\chi$ if one 
can obtain $\int \frac{d \chi}{\chi} \sqrt{\Omega}$ in 
a closed form. Thus, one has $\chi (\bar{t})$. If one 
can obtain $\int \frac{d \bar{t}}{\sqrt{\chi}}$ also in 
a closed form then one has an explicit relation between 
$\bar{t}$ and $t$. Therefore, in principle, one has 
$A (t)$ and $\chi (t)$ in a closed form. 

Evidently, solutions can be obtained thus in a closed 
form only in special cases. Even then the details of 
the required integrations and inversions of functions are 
likely to obscure the general features of the solutions. 
Such explicit solutions have been studied in 
\cite{barrow} for a few specific functions 
$\Omega (\chi)$, and in \cite{k2} for any arbitrary 
function $\Omega (\chi)$ but with $\rho = 0$. In 
the present paper, however, we follow a different 
approach which is valid for any matter and for 
any arbitrary function $\Omega (\chi)$. 

\begin{center}
{\bf 4. Analysis of the Evolution}
\end {center} 


Our main goal in this paper is to determine whether 
the evolution of the universe in the present model 
is singular or not. The task is trivial if one can solve 
the equations of motion (\ref{ad}) - (\ref{rho}) for 
arbitrary functions $\Omega (\chi)$ (or $\psi (\phi)$). 
However, explicit solutions can be obtained only in 
special cases as described above. Hence, a different 
approach is needed for the general case which is valid 
for any matter and for any arbitrary function 
$\Omega (\chi)$, 
and which enables one to analyse the evolution and obtain 
its generic features even in the absence of explicit 
solutions. 

We will present such an approach below. We analyse 
the evolution and obtain its generic features for 
the general case where matter and $\Omega (\chi)$ 
(or $\psi (\phi)$) are arbitrary. We will show, 
in particular, that when $\Omega (\chi)$ (or 
$\psi (\phi)$) satisfies the constraints in (\ref{123}) 
(or (\ref{12phi})), the evolution of the universe, such 
as our observed one, is singularity free at all times. 

To prove the presence of a singularity it suffices to 
show that any one of the curvature invariants, usually 
the curvature scalar, diverges. However, to prove 
the absence of a singularity, it is necessary to show 
that {\em all} curvature invariants are finite. As proved 
in the Appendix, a sufficient condition for {\em all} 
curvature invariants to be finite is that the quantities 
in (\ref{qty}) be all finite. Some or all of these 
quantities have a potential divergence when, for example, 
$e^A$ and/or $\chi$ vanish or when $\Omega \to \infty$. 
Hence, in the following analysis, we pay particular 
attention to these quantities, determining their 
finiteness or otherwise. We perform our analysis, as 
far as possible, in terms of $\chi$ and $\Omega (\chi)$. 
In the limit as $\chi \to 1$, however, it is sometimes 
necessary to perform the analysis in terms of $\phi$ 
and $\psi (\phi)$, as will become clear below. 

To proceed, we need initial conditions at an initial 
time, say $t = t_i$. That is, we need $\dot{A} (t_i), 
\; \dot{\chi} (t_i)$, and $\chi (t_i)$ or equivalently 
$\Omega (t_i)$. Here and in the following $\Omega (t)$ 
means $\Omega (\chi (t))$, {\em i.e.} the function 
$\Omega (\chi)$ evaluated at time $t$. Note that if 
$\dot{\chi} (t_i) = 0$ and $\chi (t_i) = 1$, then 
$\Omega (t_i) = \infty$ and the evolution is same as in 
Einstein's theory. In particular, there is a singularity 
at a finite time in the past. Therefore, we assume that 
the initial conditions in the following refer always to 
generic values of $\dot{A} (t_i)$, $\dot{\chi} (t_i)$, 
and $\Omega (t_i)$ only, {\em i.e.} $\dot{A} (t_i)$ and 
$\dot{\chi} (t_i)$ are finite and non infinitesimal 
and $\Omega (t_i) < \infty$ in the following. 

The analysis will be carried out for any arbitrary 
function $\Omega (\chi)$, satisfying only the constraints 
in (\ref{123}). Hence, only the generic features of 
the evolution can be obtained but not the numerical 
details which require choosing a specific function 
for $\Omega (\chi)$. Therefore, precise numerical 
values for $\dot{A} (t_i)$ and $\dot{\chi} (t_i)$ 
are not needed. Their signs alone are sufficient. 
We also need to know whether or not 
$\Omega (t_i) \stackrel{<}{_\sim} 3$, and also the value 
of $\epsilon$ in (\ref{ad}). Thus, there are 16 sets of 
possible initial conditions: 2 each for the signs of 
$\dot{A} (t_i), \; \dot{\chi} (t_i)$, and $\epsilon$, and 
2 for whether $\Omega (t_i) \stackrel{<}{_\sim} 3$ or not. 

However, it is not necessary to analyse all of these 
16 sets. The evolution for $- \epsilon$ is same as that 
for $\epsilon$, but with the direction of time reversed. 
Hence, only 8 sets need to be analysed. Now, let 
$\epsilon = 1$. If $\dot{\chi} < 0$ then $\dot{A}$ can 
not be negative for any value of $\Omega$, see (\ref{ad}). 
Similarly, if $\dot{\chi} > 0$ and $\Omega > 3$ then also 
$\dot{A}$ cannot be negative since 
$\sqrt{\frac{\rho}{6 \chi} 
+ \frac{\Omega \dot{\chi}^2}{12 \chi^2}}  \ge 
\sqrt{\frac{\Omega}{3}} \; \frac{\dot{\chi}}{2 \chi} 
\; > \frac{\dot{\chi}}{2 \chi}$ in equation (\ref{ad}). 
Thus, there remain only 5 sets of possible initial 
conditions. We choose these sets so as to be of direct 
relevence to the evolution of observed universe and 
analyse each of them for $t > t_i$. In the course of 
the analysis, equivalent forms of the equations of motion 
given in section 3 will become useful. We then 
use these results in sections 5 and 6 to 
describe the generic evolution of the universe. 

Two remarks are now in order. First, the amount and 
the nature of the dominant ``matter'' in the universe 
are given by $\rho_0$ and $\gamma$ where $- 1 \le \gamma 
\le 1$. The later varies as the scale factor $e^A$ of 
the universe evolves. If $e^A$ is increasing, {\em i.e.} 
if $\dot{A} > 0$, then the universe is expanding, 
eventually becoming dominated by ``matter'' with 
$\gamma < \frac{1}{3}$, when such ``matter'' is present. 
For the observed universe, which is known to contain dust 
for which $\gamma = 0$, one may take $\gamma \le 0$. Thus, 
by choosing $t_i$ suitably we may assume, with no loss of 
generality, that $\gamma < \frac{1}{3}$ ($\le 0$ for 
observed universe) if $\dot{A} > 0$. Similarly, if 
$\dot{A} < 0$ then we may assume, with no loss of 
generality, that $\gamma \ge \frac{1}{3}$ when such 
``matter'' is present. This is valid for the observed 
universe which is known to contain radiation for which 
$\gamma = \frac{1}{3}$. 

Second, consider $\sigma (t)$ in (\ref{chid}) or 
(\ref{phid}). The integrand is positive and 
$\propto \frac{1}{\sqrt{\Omega}}$. (The dependence on 
$\rho_0$ can be  absorbed into the unit of time, as 
explained in section 3.) Hence, the value of the integral 
is controlled by $\Omega$. For example, in the limit 
$\chi \to 1$, it is controlled by the constant $\Omega_1$ 
in (\ref{123}). We will use this fact in the following. 
Also, since $t > t_i$, the sign of $\sigma (t)$ is same 
as that of $(1 - 3 \gamma)$. Moreover, $\sigma (t) = 0$ 
if $\rho_0 = 0$ or $\gamma = \frac{1}{3}$ corresponding 
to an universe containing, respectively, no matter or 
radiation only. 

\begin{center}
{\bf 4 a. $\dot{A} (t_i) > 0, \; \; 
\dot{\chi} (t_i) > 0$, and $\Omega (t_i) > 3$} 
\end {center} 


These conditions imply that $\epsilon = + 1$ in 
(\ref{ad}) and the constant $c > 0$ in (\ref{chid}). 
For $t > t_i$, the scale factor $e^A$ increases and, 
eventually, $\gamma$ can be taken to be $< \frac{1}{3}$ 
when such ``matter'' is present. In fact, $\gamma \le 0$ 
for the obeserved universe. Then, $(1 - 3 \gamma) > 0$ 
and, hence, $\sigma (t)$ increases. Thus, for $t > t_i$, 
$\dot{\chi} > 0$ and $\chi$ and $\Omega$ both increase. 
Since $\Omega > 3$, it follows from equation (\ref{ad}) 
that $\dot{A} > 0$ for $t > t_i$ and, hence, $e^A$ 
increases. Eventually, $\chi \to 1$ and 
$\Omega \to \Omega_1 (1 - \chi)^{- 2 \alpha}$. 

In the limit as $\chi \to 1, \; (\sigma (t) + c) > 0$ 
is a constant to an excellent approximation. Then, 
equation (\ref{achi}) can be solved relating $\chi$ 
and $e^A$: 
\begin{equation}\label{achi1}
e^A = {\rm (constant)} \; (1 - \chi)^{- 
\frac{2 (1 - \alpha)}{3 (1 - \gamma)}} \; . 
\end{equation} 
Substituting this result in equation (\ref{chid}) then 
yields the unique solution in the limit $\chi \to 1$: 
\begin{equation}\label{soln1}
e^A = e^{A_0} t^{\frac{2}{3 (1 + \gamma)}} 
\; , \; \; \; \; 
\chi = 1 - \chi_0 t^{- \frac{1 - \gamma}
{(1 + \gamma) (1 - \alpha)}} \; ,  
\end{equation}
where $A_0$ and $\chi_0 > 0$ are constants and $t$ is 
measured in appropriate units. Since $\chi \to 1$, it 
follows from (\ref{soln1}) that $t \to \infty$. It can 
now be seen that $(\sigma (t) + c) > 0$ is indeed 
a constant to an excellent approximation. Also, 
the quantities in (\ref{qty}) are all finite, implying 
that all the curvature invariants are finite. Thus, 
there is no singularity as $\chi \to 1$ and, thus, 
for $t > t_i$. 

Note that the above relation between $\chi$ and $e^A$ 
and, hence, the solution (\ref{soln1}) is not valid if 
$\rho = 0$ or $\gamma = 1$ {\em i.e.} if the universe 
contains no matter or ``matter'' with $\gamma = 1$ only. 
Although these cases are not relevent for the observed 
universe, we will now consider them for the sake 
of completeness. The results, however, are useful 
in describing the evolution for toy universes in 
section 5. 


\begin{center}
{\bf $\rho = 0$}
\end{center}


$\rho = 0$ implies that $\rho_0 = 0$. In this case, 
$\sigma (t) = 0$ identically and the analysis was first 
given in \cite{k2}, albeit differently. The dynamics 
turns out to be clear in terms of $\phi$ and 
$\psi (\phi)$. These variables, although equivalent 
to $\chi$ and $\Omega (\chi)$, are particularly well 
suited in describing the evolution near $\phi = 0$, or 
equivalently $\chi = 1$, where $\Omega \to \infty$. 
In fact, in this limit, it is not possible sometimes 
to perform the analysis in terms of $\chi$ and 
$\Omega (\chi)$. 

At $t_i$, we have $\dot{A} > 0, \; \dot{\phi} > 0$ and, 
with our normalisation, $\phi < 0, \; \psi < 0$ and 
$\psi_{\phi} > 0$. Also, $\psi (\phi)$ has a maximum 
$\psi_{max} = 0$ at $\phi = 0$. 
The equations of motion (\ref{adpsi}) 
and (\ref{phidd}) are now given by 
\begin{equation}\label{rho0}
\dot{A} = (- \sqrt{3} \psi_{\phi} + {\rm sign} 
(\dot{\phi})) \; \frac{\dot{\phi}}{2 \sqrt{3}} \; , 
\; \; \; \; 
\dot{\phi} = c e^{- 3 A - \psi} \; , 
\end{equation}
where $c > 0$. For $t > t_i$, the scale factor $e^A$ and 
$\phi$ continue to increase. Eventually, $\phi \to 0$. 
In this limit, $\psi \to 0$ and $\psi_{\phi} \to 0$. 
The solution to (\ref{rho0}) is then given by 
\begin{equation}\label{cross}
e^A = e^{A_0} t^{\frac{1}{3}} \; , \; \; \; \; 
\phi = c \; {\rm ln} t \; , 
\end{equation}
where $t$ is measured in appropriate units. From these 
expressions, it is clear that the field $\phi$ crosses 
the value $0$ at $t = 1$ (in appropriate units) and 
becomes positive, while the scale factor $e^A$ remains 
finite and increasing. For $\phi > 0, \; \psi (\phi)$ 
decreases and $\psi_{\phi} < 0$. Equivalently, $\chi$ 
and $\Omega (\chi)$ both decrease. Thus, we now have 
$\dot{A} > 0, \; \dot{\chi} < 0$, and $\epsilon = 1$. 
Further evolution is analysed in section 4b.  

Note the crossover at $\phi = 0$ where $\phi$ crosses 
$0$ and becomes positive. In terms of $\chi$ and 
$\Omega (\chi)$, this amounts to the following: $\chi$ 
reaches $1$, correspondingly $\Omega$ reaches $\infty$, 
and then $\chi$ ``reflects'' back and decreases, 
correspondingly $\Omega$ decreases. This 
crossover/reflection takes place in a finite time and 
with $e^A$ remaining finite during the process. It is 
this phenomenon of crossover that can be seen only in 
terms of $\phi$ and $\psi (\phi)$, and not $\chi$ and 
$\Omega (\chi)$. 


\begin{center}
{\bf $\gamma = 1$} 
\end{center}


In this case, the universe contains ``matter'' with 
$\gamma = 1$ only. Hence, $\rho = \rho_0 e^{- 6 A}$ and 
$\sigma (t)$ is negative. However, as $\chi \to 1, \; 
\sigma (t) \propto \frac{1}{\sqrt{\Omega_1}}$, see 
equation (\ref{123}), and remains finite as can be 
verified a posteriori. Also, the magnitude of 
$\sigma (t)$ can be made as small as necessary by 
choosing the model dependent constant $\Omega_1$ 
sufficiently large. Hence, as $\chi \to 1$, the factor 
$(\sigma (t) + c)$ can be taken to remain positive and 
constant. 

Now, $\Omega \dot{\chi}^2 = {\rm (constant)} \; e^{- 6 A}$ 
and its dependence on $e^A$ is same as that of $\rho$.  
Therefore, in the limit $\chi \to 1$ or equivalently 
$\Omega \to \infty$, the solution to (\ref{adpsi}) and 
(\ref{phid}) and, hence, the evolution turn out to be 
identical to the $\rho = 0$ case. In particular, 
the dynamics is clear in terms of $\phi$ and 
$\psi (\phi)$ only. The field $\phi$ crosses the value 
$0$ (at $t = 1$ in appropriate units) and becomes 
positive, while the scale factor $e^A$ remains finite 
and increasing. This crossover cannot be 
seen in terms of $\chi$ and $\Omega (\chi)$.

For $\phi > 0, \; \psi (\phi)$ decreases and 
$\psi_{\phi} < 0$. Equivalently, $\chi$ and 
$\Omega (\chi)$ both decrease. Thus, we now have 
$\dot{A} > 0, \; \dot{\chi} < 0$, and $\epsilon = 1$. 
Further evolution is analysed in section 4b.  

\begin{center}
{\bf 4 b. $\dot{A} (t_i) > 0, \; \; 
\dot{\chi} (t_i) < 0, \; \; \Omega (t_i) > 3$, 
and $\epsilon = 1$} 
\end {center} 


The constant $c < 0$ in (\ref{chid}). For $t > t_i, \; 
\chi$ and $\Omega$ decrease and the scale factor $e^A$ 
increases. Equation (\ref{achi}) now becomes 
\[
2 \chi \; \frac{d A}{d \chi} = - 1 - \sqrt{K} \; , 
\]
where $K (t)$ is given in (\ref{achi}). 

Consider first the case $\rho = 0$ or $ \gamma = 1$, 
discussed in section 4a and which led to 
the present phase. Then, $\rho_0 (1 - 3 \gamma) \le 0$ 
and, therefore, $\sigma (t) \le 0$. Thus, 
$(\sigma (t) + c) \le c < 0$ and, hence, $\dot{\chi} (t)$ 
is negative and non zero, implying that $\chi (t)$ 
decreases for $t > t_i$. Also, $K (t)$ remains finite 
and it follows from equation (\ref{achi}) that 
$\frac{d A}{d \chi} < 0$. Thus, $\dot{A} > 0$ 
since $\dot{\chi} < 0$ and, hence, $e^A$ 
increases for $t > t_i$. 

Eventually, as $t$ increases, $\chi \to 0$ and 
$e^A \to \infty$. Also, $\frac{2 \rho_0 \chi 
e^{3 (1 - \gamma) A}}{(\sigma (t) + c)^2} \ll 1$ 
in (\ref{achi}) since $\rho_0 = 0$ or $\gamma = 1$. 
Equation (\ref{achi}) can then be solved relating 
$\chi$ and $e^A$: 
\begin{equation}\label{achi2}
e^A = {\rm (constant)} \; \chi^{- \frac{3 
+ \sqrt{3 \Omega_0}}{6}} \; , 
\end{equation}
where $\chi \to 0$ and $\Omega_0$ is given in 
(\ref{123}). Substituting this result in equation 
(\ref{chid}) then yields the unique solution in 
the limit $\chi \to 0$: 
\begin{equation}\label{soln2}
e^A = e^{A_0} (t - t_0)^n \; , \; \; \; \; 
\chi = \chi_0 (t - t_0)^m \; , 
\end{equation}
where $A_0, \; \chi_0$, and $t_0 > t_i$ are some 
constants and 
\begin{equation}\label{mn+}
n = \frac{3 + \sqrt{3 \Omega_0}}
{3 (1 + \sqrt{3 \Omega_0})} \; , \; \; \; \; 
m = \frac{- 2}{1 + \sqrt{3 \Omega_0}} \; . 
\end{equation}
Note that $m < 0$. Then, since $\chi \to 0$, it 
follows from (\ref{soln2}) that $t \to \infty$. 

Thus, as $\chi \to 0$, $t \to \infty$ and 
$e^A \to \infty$. It can also be seen that the quantities 
in (\ref{qty}) are all finite for $t_i \le t \le \infty$, 
implying that all the curvature invariants are finite. 
Thus, there is no singularity for $t_i \le t \le \infty$. 

Consider now the case where $\rho$ and $\gamma$ 
are non zero and have generic values. We have 
$\dot{A} (t_i) > 0$ and $\dot{\chi} (t_i) < 0$. Hence, 
$e^A$ increases and $\chi$ decreases. As $e^A$ increases, 
$\gamma$ can eventually be taken to be $< \frac{1}{3}$, 
when such ``matter'' is present. In fact, $\gamma \le 0$ 
for the obeserved universe. In such cases where 
$\gamma \le 0$, $\sigma (t)$ grows faster than $t$ as 
can be seen from (\ref{sigma}). Then, as $t$ increases, 
the factor $(\sigma (t) + c)$ becomes positive. 

This implies that $\dot{\chi}$, initially negative, 
passes through zero and becomes positive. We then have 
$\dot{A} > 0$ and $\dot{\chi} > 0$. As $t$ increases 
further, $\chi$ and, hence, $\Omega (\chi)$ increase. 
If $\Omega > 3$ then further evolution proceeds as 
described in section 4a. If $\Omega < 3$ then, 
upon making a time reversal, the initial conditions 
become identical to the one described in section 4e. 
The evolution in the present case then 
proceeds as in section 4d for the case which 
leads to the initial conditions of section 4e. 

\begin{center}
{\bf 4 c. $\dot{A} (t_i) < 0, \; \; \dot{\chi} (t_i) < 0$, 
and $\Omega (t_i) > 3$}
\end {center} 


These conditions imply that $\epsilon = - 1$ in 
(\ref{ad}) and the constant $c < 0$ in (\ref{chid}). 
For $t > t_i$, the scale factor $e^A$ decreases 
and, eventually, $\gamma$ can be taken to be 
$\ge \frac{1}{3}$ when such ``matter'' is present. In 
fact, this is the case for the observed universe. Then 
$(1 - 3 \gamma) \le 0$ and $\sigma (t)$ decreases 
or remains constant. Therefore, the factor 
$(\sigma (t) + c) \le c < 0$ and,  since $e^A$ decreases, 
we have that $\dot{\chi} (t) < \dot{\chi} (t_i) < 0$ for 
$t > t_i$. Thus $\chi$ and, hence, $\Omega$ decrease for 
$t > t_i$. Also, $\frac{d A}{d \chi} > 0$ since 
$\dot{\chi} < 0$ and $\dot {A} < 0$. 
 
In (\ref{achi}), $K (t_i) > 1$ since $\Omega > 3$. From 
the behaviour of $e^A, \; \chi, \; \Omega (\chi)$, and 
$(\sigma (t) + c)$ described above, it follows that $K$ 
decreases monotonically. The lowest value of $K$ is 
$\frac{\Omega_0}{3} \le \frac{1}{9}$, achievable when 
$\chi$ vanishes. This, together with $K (t_i) > 1$ 
and the monotonic behaviour of $K (t)$, then implies 
that there exists a time, say $t = t_m > t_i$ where 
$K (t_m) = 1$ with $\chi (t_m) > 0$. Hence, 
$\frac{d A}{d \chi} (t_m) = 0$. Therefore, 
$\dot{A} (t_m) = 0$ since $\dot{\chi} (t_m)$ 
is non zero. This is a critical point of $e^A$ and is 
a minimum. Also, equation (\ref{achi}) can be written as 
\begin{equation}\label{intachi}
A (t_i) - A (t_m) = \int^{\chi (t_i)}_{\chi (t_m)} 
\frac{d \chi}{2 \chi} \; (- 1 + \sqrt{K}) 
= {\rm finite} \; , 
\end{equation}
where the last equality follows because both 
the integrand and the interval of integration are 
finite. Therefore, $A (t_m)$ is finite and $> - \infty$ 
and, hence, $e^{A (t_m)}$ is finite and non vanishing. 
\footnote{The existence of such a non vanishing minimum 
of $e^A$ has also been deduced in the variable mass 
theories of \cite{mb}. However, singularities may 
subsequently arise in these theories. See section 7 also.} 

Thus, for $t > t_i$, the scale factor $e^A$ continues 
to decrease and reaches a non zero minimum at $t = t_m$. 
The precise values of $t_m, \; A (t_m), \; \chi (t_m)$, 
and $\Omega (t_m)$ are model dependent. Hence, nothing 
further can be said about them except, as follows from 
$K (t_m) = 1$, that $\Omega (t_m) \stackrel{<}{_\sim} 3$ 
in general and $= 3$ when $\rho_0 = 0$.  

However, the above information suffices for our purposes. 
It can now be seen that the quantities in (\ref{qty}) 
are all finite, implying that all the curvature 
invariants are finite. Thus, there is no singularity 
for $t_i \le t \le t_m$. 

For $t > t_m$, one has $\dot{A} (t) > 0, \; \dot{\chi} 
(t) < 0$, and $\Omega (\chi (t_m)) \stackrel{<}{_\sim} 
3$ by continuity. Further evolution is analysed below. 

\begin{center}
{\bf 4 d. $\dot{A} (t_i) > 0, \; \; \dot{\chi} (t_i) < 0, 
\; \; \Omega (t_i) \stackrel{<}{_\sim} 3$, and 
$\epsilon = - 1$ }
\end {center} 


This implies that the constant $c < 0$ in (\ref{chid}) 
and that $\frac{d A}{d \chi} (t_i) < 0$ in (\ref{achi}). 
Hence, $K (t_i) < 1$. For $t > t_i$, the scale factor 
$e^A$ increases and, eventually, $\gamma$ can be taken to 
be $< \frac{1}{3}$ when such ``matter'' is present. In 
fact, $\gamma \le 0$ for the observed universe. Then, 
$(1 - 3 \gamma) > 0$ and, hence, $\sigma (t)$ increases. 
Thus, $e^A$ is increases, $\chi$ decreases, and 
$(\sigma (t) + c)$ increases. Now, depending on $\rho_0$, 
$\gamma$, and the details of $\Omega (\chi)$, $K (t)$ in 
(\ref{achi}) may or may not remain $< 1$ for $t > t_i$.  

Consider first the case where $K (t)$ remains $< 1$ for 
$t > t_i$. It then follows that $(\sigma (t) + c)$ 
cannot have a zero for $t > t_i$ and, since $c < 0$, 
must remain negative and non infinitesimal. Note that 
$\gamma$ must be $> 0$ for otherwise $\sigma (t)$ grows 
as fast as or faster than $t$, making $(\sigma (t) + c)$ 
vanish at some finite time. Let 
\begin{equation}\label{ce}
\lim_{t \to \infty} (\sigma (t) + c) = c_e 
\end{equation} 
where $c_e$ is a negative, non infinitesimal constant. We 
then have for $t_i \le t \le \infty, \; c \le (\sigma (t) 
+ c) \le c_e < 0$ and, hence, $\dot{\chi} (t) < 0$. 
Therefore, eventually $\chi (t) \to 0$. Since $K (t) < 1$ 
and $\dot{\chi} (t) < 0$, it also follows from 
(\ref{achi}) that $\frac{d A}{d \chi} < 0$ and, hence, 
$\dot{A} (t) > 0$. We will now consider the limit 
$\chi \to 0$. 

$K (t) < 1$ implies that as $\chi \to 0$, 
$K \to \frac{\Omega_e}{3} < 1$ where 
\begin{equation}\label{omegae}
\Omega_e \equiv \Omega_0 \; \left( 1 
+ \frac{2 \rho_0}{c_e^2} \; \lim_{\chi \to 0} 
\chi e^{3 (1 - \gamma) A} \right) 
\end{equation} 
is a constant and $\Omega_0$ is given in (\ref{123}). 
Now, as $\chi \to 0$, equation (\ref{achi}) 
can be solved relating $\chi$ and $e^A$: 
\begin{equation}\label{achi0}
e^A = ({\rm constant}) \; 
\chi^{- \frac{3 - \sqrt{3 \Omega_e}}{6}} \; . 
\end{equation}
Note that $(3 - \sqrt{3 \Omega_e}) > 0$ since 
$\Omega_e < 3$. Hence, $e^A \to \infty$ as 
$\chi \to 0$. Substituting (\ref{achi0}) in equation 
(\ref{chid}) then yields the unique solution in 
the limit $\chi \to 0$: For $\Omega_e \ne \frac{1}{3}$, 
\begin{equation}\label{chi0}
e^A = e^{A_0} \left( t_0 - {\rm sign} (m) t \right)^n 
\; , \; \; \; \; \chi = \chi_0 
\left( t_0 - {\rm sign} (m) t \right)^m \; , 
\end{equation}
where $t$ is measured in appropriate units, $A_0, \; 
\chi_0$, and $t_0 > t_i$ are some constants, and 
\begin{equation}\label{mn} 
n = \frac{3 - \sqrt{3 \Omega_e}}
{3 (1 - \sqrt{3 \Omega_e})} \; , \; \; \; \; 
m = \frac{- 2}{1 - \sqrt{3 \Omega_e}} \; . 
\end{equation}
For $\Omega_e = \frac{1}{3}$, 
\begin{equation}\label{chi0e}
e^A = e^{A_0} e^{- \frac{c_e (t - t_0)}{3}} \; , 
\; \; \; \; \chi = \chi_0 e^{c_e (t - t_0)} \; , 
\end{equation}
where $c_e$ is defined in (\ref{ce}). 

Let $\Omega_e > \frac{1}{3}$. Hence, $m > 0$. Also, 
$n < 0$ because $\Omega_e < 3$. Since $\chi \to 0$, it 
follows from (\ref{chi0}) that $t \to t_0 > t_i$, which 
implies that $\chi$ vanishes at a finite time $t_0$. In 
this limit, the scale factor $e^A \to \infty$ since 
$n < 0$. The quantities in (\ref{qty}), for example 
$\frac{\dot{\chi}}{\chi}$, also diverge, implying that 
the curvature invariants, including the Ricci scalar, 
diverge. Thus, for $\Omega_e > \frac{1}{3}$, there is 
a singularity at a finite time $t_0$. 

Let $\Omega_e \le \frac{1}{3}$, which is the case in our 
model, see (\ref{123}). When $\Omega_e < \frac{1}{3}$, 
$m < 0$. Also, $n > 0$ because $\Omega_e < 3$. Since 
$\chi \to 0$, it follows from (\ref{chi0}) that 
$t \to \infty$. In this limit, the scale factor 
$e^A \to \infty$ since $n > 0$ now. The same result 
holds also for the solution in (\ref{chi0e}) with 
$\Omega_e = \frac{1}{3}$. 

For $K (t)$ to be $< 1$, $(\sigma (t) + c)$ must not be 
too small. Since $c < 0$ and $\sigma (t) > 0$, it 
necessarily implies that $\sigma (t)$ must remain finite. 
It follows from the above solutions that $\sigma (t)$ can 
remain finite only if $\sqrt{3 \Omega_e} > 
\frac{1 - 3 \gamma}{1 - \gamma}$, equivalently, $\gamma 
> \frac{1 - \sqrt{3 \Omega_e}}{3 - \sqrt{3 \Omega_e}}$. 
Under this condition, it can be checked easily that 
$\lim_{\chi \to 0} \chi e^{3 (1 - \gamma) A} = 0$.  
Hence, $\Omega_e = \Omega_0$ as follows from 
(\ref{omegae}). 

It can now be seen, for $\Omega_e = \Omega_0 \le 
\frac{1}{3}$, that the quantities in (\ref{qty}) are 
all finite for $t_i \le t \le \infty$, implying that 
all the curvature invariants are finite. Thus, there 
is no singularity for $t_i \le t \le \infty$ when 
$\Omega_0 \le \frac{1}{3}$.  

Consider the second case where $K (t)$ in (\ref{achi}) 
may not remain $< 1$ for all $t > t_i$. This is the case 
for the observed universe where $\gamma \le 0$ and, 
hence, $\sigma (t)$ grows as fast as or faster than $t$. 
Then there is a time, say $t = t_1$, when the value of 
$K = 1$. It follows that $c < (\sigma(t_1) + c) < 0$ and, 
hence, $\dot{\chi} (t) < 0$ for $t_i \le t \le t_1$. 
Also, $\chi (t_1)$ is non vanishing since the case of 
$\chi (t_1) \to 0$ is same as the case described above. 
Therefore, we have that $\frac{d A}{d \chi} (t_1) = 0$ 
which implies, since $\dot{\chi} (t_1) \ne 0$, that 
$\dot{A} (t_1) = 0$. This is a critical point of $e^A$ 
and is a maximum. 

For $t > t_1, K (t)$ becomes $> 1$ and, hence, 
$\dot{A} (t) < 0$. Also, $\dot{\chi} < 0$ and 
$\Omega (t) < 3$. Then $\epsilon = -1$ necessarily. 
Further evolution is analysed below. 

\begin{center}
{\bf 4 e. $\dot{A} (t_i) < 0, \; \; 
\dot{\chi} (t_i) < 0$, and $\Omega (t_i) < 3$} 
\end {center} 


The above conditions imply that $\epsilon = - 1$, 
the constant $c < 0$ in (\ref{chid}), and  
$\frac{d A}{d \chi} (t_i) > 0$. Hence, $K (t_i) > 1$. 
For $t > t_i$, $(\sigma (t) + c)$ which is negative 
at $t_i$ may or may not vanish for non zero $\chi$. 

Consider the case where $(\sigma (t) + c)$ does not 
vanish and remains negative for non zero $\chi$. 
Therefore, $\dot{\chi} (t) < 0$ and, hence, $\chi (t)$ 
decreases. For $t > t_i$, the scale factor $e^A$ 
decreases and, eventually, $\gamma$ can be taken to be 
$\ge \frac{1}{3}$ when such ``matter'' is present. In 
fact, this is the case for the observed universe. 
Then, $(1 - 3 \gamma) < 0$ and $\sigma (t)$ decreases. 
Hence, $(\sigma (t) + c)$ also decreases. 

It follows from (\ref{achi}) that $K$, initially $> 1$, 
is now decreasing since $\chi$ and $e^A$ are decreasing 
and $(\sigma (t) + c)^2$ is increasing. Its lowest value 
$= \frac{\Omega_0}{3} \le \frac{1}{9}$, achieveable at 
$\chi = 0$. Therefore, there exists a time, say 
$t = t_M$, where $K = 1$ and, hence, 
$\frac{d A}{d \chi} = 0$. Clearly, $\chi (t_M)$ and 
$\dot{\chi} (t_M)$ are non zero for the same reasons as 
in section 4c, implying that $\dot{A} (t_M) = 0$. 
This is a critical point of $e^A$ and is a minimum. 

The existence of this critical point of $e^A$ can be seen 
in another way also. As $(\sigma (t) + c) < 0$ grows in 
magnitude, it follows from equation (\ref{chid}) that 
$\Omega \dot{\chi}^2 \propto e^{- 6 A}$, whereas $\rho$ 
is given by (\ref{rho}). Note that $\dot{\chi} (t) < 0$ 
and, hence, $\chi (t)$ decreases and eventually $\to 0$. 
Then, in the limit $\chi \to 0$, 
the $\frac{\Omega \dot{\chi}^2}{\chi^2}$ term in equation 
(\ref{ad}) dominates $\frac{\rho}{\chi}$ term. Hence, in 
this limit where 
$\Omega \to \Omega_0 \le \frac{1}{3} < 3$, we have that 
$\dot{A} > 0$. Since $\dot{A} < 0$ initially, this 
implies the existence of a zero of $\dot{A}$. This is 
a critical point of $e^A$ and is a minimum. 

Note that all quantities remain finite for 
$t_i \le t \le t_M$. In particular, the quantities 
in (\ref{qty}) are all finite, implying that all 
the curvature invariants are finite. Thus, there 
is no singularity for $t_i \le t \le t_M$. 

For $t > t_M$, we have $\dot{A} > 0, \; \dot{\chi} 
< 0, \; \Omega < 3$, and $\epsilon = -1$. Further 
evolution then proceeds as described in section 4d. 
The evolution can thus become repetetive and oscillatory. 

Consider now the case where $(\sigma (t) + c)$ vanishes,
say at $t = t_{m'}$, for non zero $\chi$. Hence, 
$\dot{\chi} (t_{m'})$ vanishes and this is a minimum 
of $\chi$. For $t > t_{m'}, \; (\sigma (t) + c)$ and 
$\dot{\chi}$ are positive by continuity and, hence, 
$\chi$ increases. Also, $\dot{A} < 0$ and the scale 
factor $e^A$ decreases. Eventually, $\gamma$ can be 
taken to be $\ge \frac{1}{3}$ when such ``matter'' is 
present. In fact, this is the case for the observed 
universe. Then, $(1 - 3 \gamma) < 0$ and $\sigma (t)$ 
decreases. Hence, $(\sigma (t) + c)$ also decreases. 

Now, as $\chi$ increases and $\to 1$, $(\sigma (t) + c)$ 
may or may not vanish with $\chi < 1$.  If 
$(\sigma (t) + c)$ vanishes with $\chi < 1$, then 
$\dot{\chi}$ vanishes. This is a critical point of $\chi$ 
and is a maximum, beyond which we have $\dot{A} < 0, \; 
\dot{\chi} < 0$, and $\epsilon = -1$. Further evolution 
then proceeds as described in section 4c 
if $\Omega > 3$, and as in section 4e if 
$\Omega < 3$. The evolution can thus become repetetive 
and oscillatory.  

If $(\sigma (t) + c)$ does not vanish and remain 
positive for $\chi \le 1$ then, eventually,
$\chi \to 1$. It follows from equation (\ref{chid}) 
that $\Omega \dot{\chi}^2 \propto e^{- 6 A}$, 
whereas $\rho$ is given by (\ref{rho}). Note that 
$\dot{A} < 0$ and $e^A$ is decreasing. Then, in this 
limit, the $\frac{\Omega \dot{\chi}^2}{\chi^2}$ term in 
equation (\ref{ad}) dominates or, if $\gamma = 1$, is 
of the same order as the $\frac{\rho}{\chi}$ term. 

Thus we have $\dot{A} < 0, \; \dot{\chi} > 0, \; \chi 
\to 1$, equivalently $\Omega \to \infty$, and $\epsilon 
= - 1$. Also, $\frac{\Omega \dot{\chi}^2}{\chi^2}$ term 
in equation (\ref{ad}) dominates or, if $\gamma = 1$, is 
of the same order as the $\frac{\rho}{\chi}$ term. 
But this is precisely the time reversed version 
of the evolution analysed in section 4b and 
in section 4a for the case which led to 
the initial conditions of section 4a, where 
the dynamics is clear in terms of $\phi$ and 
$\psi (\phi)$. Applying the results of sections 4a 
and 4b, it follows that $\phi$ will cross 
the value $0$ after which $e^{\psi}$ and, hence, $\chi$ 
begins to decrease. We then have $\dot{A} < 0$, 
$\dot{\chi} < 0$, and $\Omega > 3$. Also, 
$\epsilon = - 1$ necessarily. Further evolution then 
proceeds as described in section 4c. The evolution 
can thus become repetetive and oscillatory.

\begin{center}
{\bf 5. Evolution of toy Universes} 
\end {center} 


We now use the results of the analysis in section 4 
to describe the generic evolution of universe in 
the present model. For the purpose of illustration we 
first describe in this section the generic evolution of 
three toy universes: One, where the universe contains no 
matter, {\em i.e.} $\rho = 0$. Two, where the universe 
contains radiation only, {\em i.e.} $\rho \ne 0$ but 
$\gamma = \frac{1}{3}$. And three, where the universe 
contains massless scalar field only, {\em i.e.} 
$\rho \ne 0$ but $\gamma = 1$. Clearly, however, none 
of these toy models are realistic. The generic evolution 
of a realistic universe, such as our observed one, will 
be described in the next section. 

Note that $\sigma (t) = 0$ in the first two models. 
As described in section 3, it may then be possible 
to obtain closed form solutions, although only in 
special cases \cite{barrow,k2}. Moreover, the details of 
the specific calculations obscure the general features 
of the solutions. Hence, and also for the purpose of 
illustration, we describe the evolution of the toy 
universes within the framework of our analysis. However, 
the explicit evolutions described in \cite{barrow,k2} 
for specific cases all conform to the ones described 
by the present general analysis. 

We first describe the evolution for $t > t_i$ taking, as 
initial conditions $\dot{A} (t_i) > 0, \; \dot{\chi} 
(t_i) > 0$, and $\Omega (t_i) > 3$. Then, $\epsilon = 1$ 
necessarily. To describe the evolution for $t < t_i$, 
we reverse the direction of time and take, as initial 
conditions, $\dot{A} (t_i) < 0, \; \dot{\chi} (t_i) < 0$, 
and $\Omega (t_i) > 3$. Then, $\epsilon = - 1$ 
necessarily. The required evolution is that for $t > t_i$ 
in terms of the reversed time variable, which is also 
denoted as $t$. The results of section 4 can then 
be applied directly. 

When the initial conditions are different, the evolution 
can again be analysed along similar lines as follows. 
However, the main result that the evolution is 
singularity free remains unchanged. 

\begin{center}
{\bf 5 a. $\rho = 0, \; \; \; t > t_i$} 
\end{center} 


The evolution for this case was first described in 
\cite{k2} where explicit solutions to the equations 
of motion were obtained for any arbitrary function 
$\Omega (\chi)$ in terms of the variables defined in 
(\ref{tanew}). We will now describe this evolution 
within the framework of the analysis presented in 
section 4 which is valid for any matter and 
for any arbitrary function $\Omega (\chi)$. 

Initially, we have $\dot{A} (t_i) > 0, \; \dot{\chi} 
(t_i) > 0$, and $\Omega (t_i) > 3$. Then, $\epsilon = 1$ 
necessarily. For $t > t_i$, the evolution proceeds as 
described in section 4a. Both $e^A$ and $\chi$ 
increase. Hence, $\Omega$ increases. Eventually, 
$\chi \to 1$ and $\Omega \to \infty$. The dynamics near 
$\chi = 1$, equivalently $\phi = 0$ in our normalisation, 
is clear in terms of $\phi$ and $\psi (\phi)$. It 
follows from the analysis of section 4a that 
$\phi$ crosses $0$ and continues to increase. Then, 
$\psi (\phi)$ and, hence, $\Omega$ decrease. 

We then have $\dot{A} > 0$, but $\dot{\chi} < 0$. 
The evolution then proceeds as described in section 4b. 
$e^A$ increases whereas $\chi$ and, hence, 
$\Omega$ decrease. As $t \to \infty, \; e^A$ and $\chi$ 
evolve as given in (\ref{soln2}) and (\ref{mn+}). 

In particular, it can be seen that the quantities in 
(\ref{qty}) are all finite for $t_i \le t \le \infty$, 
implying that all the curvature invariants are finite. 
Hence, the evolution is singularity free. 

\begin{flushleft}
{\bf $t < t_i$, equivalently $(- t) > (- t_i)$} 
\end{flushleft} 


To describe the evolution for $t < t_i$, we reverse 
the direction of time. Then, in terms of the reversed 
time variable also denoted as $t$, we have 
$\dot{A} (t_i) < 0$, $\dot{\chi} (t_i) < 0$, and 
$\Omega (t_i) > 3$. Then, $\epsilon = - 1$ necessarily. 
For $t > t_i$, the evolution proceeds as described in 
section 4c. Both $e^A$ and $\chi$ decrease. Hence, 
$\Omega$ decreases. Then, as shown in section 4c, 
there exists a time, say $t = t_m > t_i$ where 
$K (t_m) = 1$ in equation (\ref{achi}), 
$e^{A (t_m)} > 0$, $\chi (t_m) > 0$, and 
$\dot{\chi} (t_m) < 0$. Note that, in this 
case where $\rho$ and, hence, $\rho_0$ vanish, 
we have $\Omega (t_m) = 3$. 

As shown in section 4c, $K (t_m) = 1$ implies 
that $\dot{A} (t_m) = 0$. Hence, the scale factor $e^A$ 
reaches a minimum. For $t > t_m$, we then have 
$\dot{A} > 0$, $\dot{\chi} < 0$, $\Omega < 3$, and 
$\epsilon = - 1$. The evolution proceeds as described 
in section 4d. 

It follows that $\sigma (t) = 0$ since $\rho$ and, 
hence, $\rho_0$ vanishes. Hence, in equation (\ref{ce}), 
$c_e = c$ which is negative and non infinitesimal. 
Therefore, as described in section 4d, $e^A$ 
increases whereas $\chi$ and, hence, $\Omega$ decrease. 
As $t \to \infty$, $e^A$ and $\chi$ evolve as given 
in (\ref{chi0}) - (\ref{chi0e}). 

In particular, it can be seen that, for $\Omega_0 \le 
\frac{1}{3}$, the quantities in (\ref{qty}) are 
all finite for $t_i \le t \le \infty$, implying 
that all the curvature invariants are finite. 
Hence, the evolution is singularity free. 

Thus, the evolution of the toy universe with $\rho = 0$ 
and with the initial conditions $\dot{A} (t_i) > 0$, 
$\dot{\chi} (t_i) > 0$, and $\Omega (t_i) > 3$ proceeds 
in the present model as follows. For $t > t_i$, the scale 
factor $e^A$ increases continuously to $\infty$. 
The field $\chi$ increases, reaches the value $1$, and 
then decreases continuously to $0$. Correspondingly, 
$\Omega$ increases, reaches $\infty$, and then decreases 
continuously to $\Omega_0$, given in (\ref{123}).  

For $t < t_i$, the scale factor $e^A$ decreases, reaches 
a non zero minimum, and then increases continuously to 
$\infty$. The field $\chi$ decreases continuously to $0$. 
Correspondingly, $\Omega$ decreases continuously to 
$\Omega_0$, given in (\ref{123}).  

Also, the curvature invariants all remain finite for 
$- \infty \le t \le \infty$. Hence, in the present model, 
the evolution of the toy universe with $\rho = 0$ is 
singularity free. 

When the initial conditions are different, the evolution 
can again be analysed along similar lines. However, 
the main result that the evolution is singularity free 
remains unchanged. 

\begin{center}
{\bf 5 b. $\gamma = \frac{1}{3}, \; \; t > t_i$} 
\end {center} 


The evolution for this case was first described in 
\cite{barrow} where explicit solutions to the equations 
of motion were obtained for a few specific functions 
$\Omega (\chi)$. We will now describe this evolution 
within the framework of the analysis presented in 
section 4 which is valid for any matter and 
for any arbitrary function $\Omega (\chi)$. 

Initially, we have $\dot{A} (t_i) > 0, \; \dot{\chi} 
(t_i) > 0$, and $\Omega (t_i) > 3$. Then, $\epsilon = 1$ 
necessarily. For $t > t_i$, the evolution proceeds as 
described in section 4a. Both $e^A$ and $\chi$ 
increase. Hence, $\Omega$ increases. As $t \to \infty$, 
$e^A$ and $\chi$ evolve as given in (\ref{soln1}) where 
$\gamma = \frac{1}{3}$ now. 

In particular, it can be seen that the quantities in 
(\ref{qty}) are all finite for $t_i \le t \le \infty$, 
implying that all the curvature invariants are finite. 
Hence, the evolution is singularity free. 

\begin{flushleft}
{\bf $t < t_i$, equivalently $(- t) > (- t_i)$} 
\end{flushleft} 


To describe the evolution for $t < t_i$, we reverse 
the direction of time. Then, in terms of the reversed 
time variable also denoted as $t$, we have 
$\dot{A} (t_i) < 0$, $\dot{\chi} (t_i) < 0$, and 
$\Omega (t_i) > 3$. Then, $\epsilon = - 1$ necessarily. 
For $t > t_i$, the evolution proceeds as described 
in section 4c. Both $e^A$ and $\chi$ decrease. 
Hence, $\Omega$ decreases. Then, as shown in section 4c, 
there exists a time, say $t = t_m > t_i$ where 
$K (t_m) = 1$ in equation (\ref{achi}), 
$e^{A (t_m)} > 0$, $\chi (t_m) > 0$, and 
$\dot{\chi} (t_m) < 0$. Note that $\Omega (t_m) < 3$. 

As shown in section 4c, $K (t_m) = 1$ 
implies that $\dot{A} (t_m) = 0$. Hence, the scale 
factor $e^A$ reaches a minimum. For $t > t_m$, we then 
have $\dot{A} > 0$, $\dot{\chi} < 0$, $\Omega < 3$, and 
$\epsilon = - 1$. The evolution proceeds as described 
in section 4d. 

It follows that $\sigma (t) = 0$ since 
$\gamma = \frac{1}{3}$. Hence, in equation (\ref{ce}), 
$c_e = c$ which is negative and non infinitesimal. 
Therefore, as described in section 4d, $e^A$ 
increases whereas $\chi$ and, hence, $\Omega$ decrease. 
As $t \to \infty, \; e^A$ and $\chi$ evolve as given 
in (\ref{chi0}) - (\ref{chi0e}). 

In particular, it can be seen that, for $\Omega_0 \le 
\frac{1}{3}$, the quantities in (\ref{qty}) 
are all finite for $t_i \le t \le \infty$, implying 
that all the curvature invariants are finite. Hence, 
the evolution is singularity free. 

Thus, the evolution of the toy universe containing 
{\em only} radiation, {\em i.e.} $\rho \ne 0$ but 
$\gamma = \frac{1}{3}$, and with the initial conditions 
$\dot{A} (t_i) > 0, \; \dot{\chi} (t_i) > 0$, and 
$\Omega (t_i) > 3$ proceeds in the present model as 
follows. For $t > t_i$, the scale factor $e^A$ increases 
continuously to $\infty$. The field $\chi$ increases 
continuously to $1$. Correspondingly, $\Omega$ 
increases continuously to $\infty$. 

For $t < t_i$, the scale factor $e^A$ decreases, reaches 
a non zero minimum, and then increases continuously to 
$\infty$. The field $\chi$ decreases continuously to $0$. 
Correspondingly, $\Omega$ decreases continuosly to 
$\Omega_0$, given in (\ref{123}).  

Also, the curvature invariants all remain finite for 
$- \infty \le t \le \infty$. Hence, in the present model, 
the evolution of the toy universe with $\rho \ne 0$ 
but $\gamma = \frac{1}{3}$ is singularity free. 

When the initial conditions are different, the evolution 
can again be analysed along similar lines. However, 
the main result that the evolution is singularity free 
remains unchanged. 

\begin{center}
{\bf 5 c. $\gamma = 1, \; \; t > t_i$} 
\end{center} 


Initially, we have $\dot{A} (t_i) > 0, \; \dot{\chi} 
(t_i) > 0$, and $\Omega (t_i) > 3$. Then, $\epsilon = 1$ 
necessarily. For $t > t_i$, the evolution proceeds as 
described in section 4a. Both $e^A$ and $\chi$ 
increase. Hence, $\Omega$ increases. 

The constant $c > 0$ in (\ref{chid}). However, 
$(1 - 3 \gamma) < 0$ and, hence, $\sigma (t) < 0$. 
Therefore, $(\sigma (t) + c)$ can become negative for 
$\chi < 1$, or remain positive as $\chi \to 1$. In 
the case where $(\sigma (t) + c)$ becomes negative 
for $\chi < 1$, it must pass through a zero where 
$\dot{\chi}$ vanishes. This critical point is a maximum 
of $\chi (t)$ beyond which we have $\dot{A} > 0$ and 
$\dot{\chi} < 0$, and $\Omega > 3$. 

Consider the case where $(\sigma (t) + c)$ remains 
positive as $\chi \to 1$. Furthermore, as described in 
section 4, it can be taken to be non infinitesimal 
by choosing the model dependent constant $\Omega_1$ in 
(\ref{123}) large enough. The dynamics near $\chi = 1$, 
equivalently $\phi = 0$ in our normalisation, is clear 
in terms of $\phi$ and $\psi (\phi)$. It follows from 
the analysis of section 4a that $\phi$ crosses 
$0$ and continues to increase. Then, $\psi (\phi)$ and, 
hence, $\Omega$ decrease. We then have $\dot{A} > 0$, 
but $\dot{\chi} < 0$. 

Thus, in both of the above cases, we now have $\dot{A} 
> 0$ and $\dot{\chi} < 0$. The evolution then proceeds 
as described in section 4b. $e^A$ increases 
whereas $\chi$ and, hence, $\Omega$ decrease. As 
$t \to \infty$, $e^A$ and $\chi$ evolve as given in 
(\ref{soln2}) and (\ref{mn+}). 

In particular, it can be seen that the quantities in 
(\ref{qty}) are all finite for $t_i \le t \le \infty$, 
implying that all the curvature invariants are finite. 
Hence, the evolution is singularity free. 

\begin{flushleft}
{\bf $t < t_i$, equivalently $(- t) > (- t_i)$} 
\end{flushleft} 


To describe the evolution for $t < t_i$, we reverse 
the direction of time. Then, in terms of the reversed 
time variable which is also denoted as $t$, we have 
$\dot{A} (t_i) < 0$, $\dot{\chi} (t_i) < 0$, and 
$\Omega (t_i) > 3$. Then, $\epsilon = - 1$ necessarily. 
For $t > t_i$, the evolution proceeds as described in 
section 4c. Both $e^A$ and $\chi$ decrease. Hence, 
$\Omega$ decreases. Then, as shown in section 4c, 
there exists a time, say $t = t_m > t_i$ where 
$K (t_m) = 1$ in equation (\ref{achi}), 
$e^{A (t_m)} > 0$, $\chi (t_m) > 0$, and 
$\dot{\chi} (t_m) < 0$. Note that $\Omega (t_m) < 3$. 

As shown in section 4c, $K (t_m) = 1$ 
implies that $\dot{A} (t_m) = 0$. Hence, the scale 
factor $e^A$ reaches a minimum. For $t > t_m$, we then 
have $\dot{A} > 0$, $\dot{\chi} < 0$, $\Omega < 3$, and 
$\epsilon = - 1$. The evolution proceeds as described 
in section 4d. 

It follows that $\sigma (t) = 0$ since $\gamma = 1$. 
Note that the constant $c < 0$ in (\ref{chid}). Thus, 
$(\sigma (t) + c) < c < 0$ and, hence, $\dot{\chi} < 0$ 
for $t > t_i$. Therefore, $\chi$ decreases. Eventually, 
$\chi \to 0$. Moreover, and as can be verified 
a posteriori, $\sigma (t) \to {\rm (constant)}$ as 
$\chi \to 0$. Hence, in equation (\ref{ce}), the constant 
$c_e$ is negative and non infinitesimal. Therefore, as 
described in section 4d, $e^A$ increases whereas 
$\chi$ and, hence, $\Omega$ decrease. As $t \to \infty$, 
$e^A$ and $\chi$ evolve as given in 
(\ref{chi0}) - (\ref{chi0e}). 

In particular, it can be seen that, for $\Omega_0 \le 
\frac{1}{3}$, the quantities in (\ref{qty}) 
are all finite for $t_i \le t \le \infty$, implying 
that all the curvature invariants are finite. Hence, 
the evolution is singularity free. 

Thus, the evolution of the toy universe containing 
{\em only} massless scalar field, {\em i.e.} $\rho \ne 0$ 
but $\gamma = 1$, and with the initial conditions 
$\dot{A} (t_i) > 0$, $\dot{\chi} (t_i) > 0$, and 
$\Omega (t_i) > 3$ proceeds in the present model as 
follows. For $t > t_i$, the scale factor $e^A$ increases 
continuously to $\infty$. The field $\chi$ increases, 
reaches a maximum $\le 1$, and then decreases 
continuously to $0$. Correspondingly, $\Omega$ increases, 
reaches a maximum $\le \infty$, and then decreases 
continuously to $\Omega_0$, given in (\ref{123}).  

For $t < t_i$, the scale factor $e^A$ decreases, reaches 
a non zero minimum, and then increases continuously to 
$\infty$. The field $\chi$ decreases continuously to $0$. 
Correspondingly, $\Omega$ decreases continuously to 
$\Omega_0$, given in (\ref{123}).  

Also, the curvature invariants all remain finite for 
$- \infty \le t \le \infty$. Hence, in the present model, 
the evolution of the toy universe with $\rho \ne 0$ 
but $\gamma = 1$ is singularity free. 

When the initial conditions are different, the evolution 
can again be analysed along similar lines. However, 
the main result that the evolution is singularity free 
remains unchanged. 

\begin{center}
{\bf 6. Evolution of observed Universe} 
\end {center} 


We now use the results of the analysis in section 4 
to describe the generic evolution of a realistic 
universe, such as our observed one, in the present model. 
A realistic universe may contain different types of 
``matter'' described by different values of $\gamma$, but 
with $- 1 \le \gamma \le 1$. For example, $\gamma = - 1, 
\; 0, \; \frac{1}{3}$, and $1$ respectively for vacuum 
energy, dust, radiation, and massless scalar field.
Our observed universe is certainly known to contain dust 
and radiation. However, in a variety of models, vacuum 
energy and/or massless scalar field are often assumed to 
be present. Our assumption that $- 1 \le \gamma \le 1$ 
then covers all these cases. As follows from (\ref{rho}), 
``matter'' with larger $\gamma$ dominates the evolution 
for smaller value of $e^A$ and vice versa. Therefore, 
when $e^A$ is decreasing we take $\gamma \ge \frac{1}{3}$ 
eventually, and when $e^A$ is increasing we take 
$\gamma \le 0$ eventually. 

In considering the evolution of the universe, we 
start with an initial time $t_i$, corresponding to 
a temperature, say $\stackrel{>}{_\sim} 10^{16}$ GeV,  
such that GUT symmetry breaking, inflation, and other 
(matter) model dependent phenomena may occur for 
$t > t_i$ only. Our observed universe is expanding at 
$t_i$ and, hence, $\dot{A} (t_i) > 0$. For the sake of 
definiteness, we take $\dot{\chi} (t_i) > 0$ and 
$\Omega (t_i) > 3$, equivalently $\omega (t_i) > 0$, 
as commonly assumed. 

We first describe the evolution for $t > t_i$ taking, as 
initial conditions $\dot{A} (t_i) > 0, \; \dot{\chi} 
(t_i) > 0$, and $\Omega (t_i) > 3$. Then, $\epsilon = 1$ 
necessarily. To describe the evolution for $t < t_i$, 
we reverse the direction of time and take, as initial 
conditions, $\dot{A} (t_i) < 0, \; \dot{\chi} (t_i) < 0$, 
and $\Omega (t_i) > 3$. Then, $\epsilon = - 1$ 
necessarily. The required evolution is that for $t > t_i$ 
in terms of the reversed time variable, which is also 
denoted as $t$. The results of section 4 can then 
be applied directly. 

When the initial conditions are different, the evolution 
can again be analysed along similar lines as follows. 
However, the main result that the evolution is 
singularity free remains unchanged. 

\begin{center}
{\bf $t > t_i$} 
\end{center} 


Initially, we have $\dot{A} (t_i) > 0, \; \dot{\chi} 
(t_i) > 0$, and $\Omega (t_i) > 3$. Then, $\epsilon = 1$ 
necessarily. For $t > t_i$, the evolution proceeds as 
described in section 4a. Both $e^A$ and $\chi$ 
increase. Hence, $\Omega$ increases. Eventually, as 
$e^A$ increases, we can take $\gamma \le 0$. Then, 
as $t \to \infty$, $e^A$ and $\chi$ evolve as given 
in (\ref{soln1}). 

In particular, it can be seen that the quantities in 
(\ref{qty}) are all finite for $t_i \le t \le \infty$, 
implying that all the curvature invariants are finite. 
Hence, the evolution is singularity free. 

Thus, in the present day universe in our model, 
$\chi (today) \to 1$ and $\Omega (today) \to \infty$. 
Also, as follows from (\ref{123}), 
$\frac{1}{\Omega^3} \frac{d \Omega}{d \chi} (today) 
= - \frac{2 \alpha}{\Omega_1^2} 
(1 - \chi)^{4 \alpha - 1} \to 0$. Therefore, 
our model satisfies the observational constraints 
imposed by solar system experiments, namely 
$\Omega (today) > 2000$ and $\frac{1}{\Omega^3} 
\frac{d \Omega}{d \chi} (today) < 0.0002$ 
(see \cite{will} pg. 117, 124-5, and 339). 

\begin{center}
{\bf $t < t_i$, equivalently $(- t) > (- t_i)$} 
\end{center} 


To describe the evolution for $t < t_i$, we reverse 
the direction of time. Then, in terms of the reversed 
time variable, also denoted as $t$, we have 
$\dot{A} (t_i) < 0$, $\dot{\chi} (t_i) < 0$, and 
$\Omega (t_i) > 3$. Then, $\epsilon = - 1$ necessarily. 
For $t > t_i$, the evolution proceeds as described in 
section 4c. Both $e^A$ and $\chi$ decrease. Hence, 
$\Omega$ decreases. Then, as shown in section 4c, 
there exists a time, say $t = t_m > t_i$ where 
$K (t_m) = 1$ in equation (\ref{achi}), 
$e^{A (t_m)} > 0$, $\chi (t_m) > 0$, and 
$\dot{\chi} (t_m) < 0$. Note that $\Omega (t_m) < 3$. 

As shown in section 4c, $K (t_m) = 1$ 
implies that $\dot{A} (t_m) = 0$. Hence, the scale 
factor $e^A$ reaches a minimum. For $t > t_m$, we then 
have $\dot{A} > 0$, $\dot{\chi} < 0$, $\Omega < 3$, and 
$\epsilon = - 1$. The evolution proceeds as described in 
section 4d. Now, however, the evolution for 
$t > t_m$ is complicated since the universe is known to 
contain ``matter'' with $\gamma = 0$. Nevertheless, all 
qualitative features of its evolution can be obtained 
using the results of section 4. 

For $t \stackrel{>}{_\sim} t_m$, $\dot{A} > 0$, 
$\dot{\chi} < 0$, and $(\sigma (t) + c) < 0$. Hence, 
$e^A$ increases and $\chi$ decreases and, eventually, 
``matter'' with $\gamma \le 0$ dominates the evolution. 
Then $(1 - 3 \gamma) > 0$ and $\sigma (t)$ begins to 
increase. Hence $(\sigma (t) + c)$, which is negative, 
also begins to increase. For $\gamma \le 0$, $\sigma (t)$ 
grows atleast as fast as $t$, as follows from 
(\ref{sigma}). Then eventually, as described in section 
4d, $e^A$ reaches a maximum eventually at time, 
say $t = t_1$. Also, $(\sigma (t_1) + c) < 0$ and, hence, 
$\dot{\chi} (t_1) < 0$. Note that this critical point of 
$e^A$ exists independent of the details of the model, 
as long as ``matter'' with $\gamma \le 0$ is present. 
Further  evolution then proceeds as described in section  
4e. 

For $t \stackrel{>}{_\sim} t_1$, we then have $\dot{A} 
< 0, \; \dot{\chi} < 0$, and $\Omega < 3$. Also, 
$(\sigma (t) + c) < 0$ but $\sigma (t)$ remains 
increasing. Hence, depending now on the details of 
the model, $(\sigma (t) + c)$ may remain negative for all 
$t > t_1$, or it may reach a zero and become positive. 

In the first case, $(\sigma (t) + c)$ remains negative 
for all $t > t_1$. Hence, $\dot{\chi} < 0$ and $\chi$ 
continues to decrease. The scale factor $e^A$, as shown 
in section 4e, reaches a minimum at time, say 
$t = t_M > t_1$, with $\chi (t_M) > 0$ non vanishing. 
For $t > t_M, \; e^A$ increases, and we have $\dot{A} > 0, 
\; \dot{\chi} < 0$, and $\Omega < 3$. Further evolution 
then proceeds as described in section 4d. 

Note that this means that the scale factor increases and 
reaches a maximum, then decreases and reaches a minimum, 
then increases and so on. The existence of maxima of 
$e^A$ is model independent as long as ``matter'' with 
$\gamma \le 0$ is present, which is the case for 
the observed universe. The minima of $e^A$ are all non 
zero for the reasons described in section 4c. 
Their existence depends on whether $(\sigma (t) + c)$ 
remains negative or not, but is otherwise model 
independent as long as ``matter'' with 
$\gamma \ge \frac{1}{3}$ is present, which is the case 
for the observed universe. The evolution can thus become 
repetetive and oscillatory. 

In the second case, $(\sigma (t) + c)$ reaches a zero 
at time, say $t = t_{m'} > t_1$ and becomes positive. 
It then follows that $\dot{\chi} (t_{m'}) = 0$ and that 
this is a minimum of $\chi$. For $t > t_{m'}$, we have 
$\dot{A} < 0$, $\dot{\chi} > 0$, and $\Omega < 3$. 
Hence, $e^A$ decreases and $\chi$ increases. 
Eventually, we can take $\gamma \ge \frac{1}{3}$. 

Now, $(1 - 3 \gamma) \le 0$ and $\sigma (t)$ begins to 
decrease or remains constant. Hence $(\sigma (t) + c)$, 
which is positive, also begins to decrease or remains 
constant. Thus, depending on the details of $\rho_0$ 
and $\gamma$, $(\sigma (t) + c)$ may or may not vanish 
with $\chi < 1$. If $(\sigma (t) + c)$ vanishes with 
$\chi < 1$, then $\dot{\chi}$ vanishes. This critical 
point is a maximum of $\chi$, beyond which we have 
$\dot{A} < 0$, $\dot{\chi} < 0$, and $\epsilon = -1$. 
Further evolution then proceeds as described in section 4c 
if $\Omega > 3$ and as in section 4e if 
$\Omega < 3$. The evolution can thus become repetetive 
and oscillatory. 

If $(\sigma (t) + c)$ does not vanish and remain 
positive for $\chi < 1$ then, eventually, $\chi \to 1$. 
It follows from equation (\ref{chid}) that 
$\Omega \dot{\chi}^2 \propto e^{- 6 A}$, whereas $\rho$ 
is given by (\ref{rho}). Note that $\dot{A} < 0$ and 
$e^A$ is decreasing. Then, in this limit, 
the $\frac{\Omega \dot{\chi}^2}{\chi^2}$ term in 
equation (\ref{ad}) dominates or, if $\gamma = 1$, 
is of the order of the $\frac{\rho}{\chi}$ term. 

Thus we have $\dot{A} < 0$, $\dot{\chi} > 0$, 
$\chi \to 1$, equivalently $\Omega \to \infty$, 
and $\epsilon = - 1$. Also, 
$\frac{\Omega \dot{\chi}^2}{\chi^2}$ term in equation 
(\ref{ad}) dominates or is of the order of 
the $\frac{\rho}{\chi}$ term. But this is precisely 
the time reversed version of the evolution analysed in 
section 4b, where the dynamics is clear in terms 
of $\phi$ and $\psi (\phi)$. Applying the results of 
section 4b, it then follows that $\phi$ will cross 
the value $0$ after which $e^{\psi}$ and, hence, $\chi$ 
will begin to decrease. We then have $\dot{A} < 0$, 
$\dot{\chi} < 0, \Omega > 3$. The evolution then 
proceeds as described in section 4c. The 
evolution can thus become repetetive and oscillatory. 

Thus it is clear that depending on the details of 
the model, the universe undergoes oscillations, perhaps 
infinitely many. As follows from the above description, 
the oscillations can stop, if at all, only in the limit 
$\chi \to 0$. The solutions then are given by 
(\ref{chi0}) - (\ref{chi0e}). In particular, however, 
the quantities in (\ref{qty}) all remain finite during 
the oscillations, implying that all the curvature 
invariants remain finite. They also remain finite 
in the solutions (\ref{chi0}) - (\ref{chi0e}) 
if $\Omega_0 \le \frac{1}{3}$ in (\ref{123}). 

Thus, it follows that if $\Omega_0 \le \frac{1}{3}$ 
then the quantities in (\ref{qty}) are all finite for 
$t_i \le t \le \infty$, implying that all the curvature 
invariants are finite. Hence, the evolution is 
singularity free. 

In summary, the evolution of a realistic universe, such 
as our observed one, with the initial conditions 
$\dot{A} (t_i) > 0$, $\dot{\chi} (t_i) > 0$, and 
$\Omega (t_i) > 3$ proceeds in the present model as 
follows. For $t > t_i$, the scale factor $e^A$ increases 
continuously to $\infty$. The field $\chi$ increases 
continuously to $1$. Correspondingly, $\Omega$ increases 
continuously to $\infty$. 

For $t < t_i$, the scale factor $e^A$ decreases and 
reaches a non zero minimum. It then increases and 
reaches a maximum, then decreases and reaches a minimum, 
then increases and so on, perhaps ad infintum. 
The oscillations can stop, if at all, only in the limit 
$\chi \to 0$. The solutions then are given by 
(\ref{chi0}) - (\ref{chi0e}). The field $\chi$ 
may, depending on the details of model, continuously 
decrease to $0$, or undergo oscillations, its maxima 
always being $\le 1$. 

Also, the curvature invariants all remain finite for 
$- \infty \le t \le \infty$. Hence, the evolution, 
although complicated and model dependent in details, 
is completely singularity free. Thus we have that 
a homogeneous isotropic universe, such as our observed 
one, evolves in the present model with no big bang or 
any other singularity. The time continues indefinitely 
into the past and the future, without encountering any 
singularity. 

When the initial conditions are different, the evolution 
can again be analysed along similar lines. However, 
the main result that the evolution is singularity free 
remains unchanged. 

\begin{center}
{\bf 7. Generalisations of the model} 
\end {center} 


The analysis and the results of sections 4 - 6
are generic and valid for any function 
$\Omega$ satisfying only the constraints (\ref{123}). 
However, these constraints can be relaxed further 
thus generalising the model, but still maintaining 
the evolution singularity free. We now study 
these generalisations. 

We first summarise the essential features of 
the constraints (\ref{123}) on $\Omega (\chi)$ by noting 
the following points. First, as evident from the results 
of the previous sections, any singularity is likely to 
arise, if at all, only in the limit $\chi \to 0$. It 
then follows from the relevent solutions (\ref{chi0}) 
- (\ref{chi0e}) that this potential singularity is 
avoided by the constraint $\Omega_0 \le \frac{1}{3}$ 
in (\ref{123}). 

Second, we used only the sufficiency condition given in 
the Appendix to determine the absence of singularity. 
Hence, the requirement of finiteness of $\frac{d^n 
\Omega}{d \chi^n}, \; \forall n \ge 1$. 

Third, there is no singularity in the limit 
$\chi \to 1$, although $\Omega \to \infty$. 
The evolution of the present day universe corresponds 
to this limit. Also, the behaviour of $\Omega (\chi)$ 
in this limit, in particular the constraint on $\alpha$ 
given in (\ref{123}) which follows naturally from 
(\ref{12phi}), ensures that our model satisfies 
the observational constraints imposed by solar system 
experiments. Note, however, that the observational 
constraints are satisfied for any $\alpha > \frac{1}{4}$. 

Fourth, the monotonicity of  $\Omega (\chi)$ for $0 < \chi 
< 1$ is not required either to avoid the singularity or 
to conform with the observations. It was invoked only to 
keep the analysis simple and transparent. Without this 
requirement, the evolution would be more complicated, its 
history possibly bearing the imprints of the ups and 
downs of $\Omega (\chi)$. The analysis in section 4 
would then involve further subclasses not 
directly relevent for the issue of singularity. 

It is now clear how the present model, equivalently 
the constraints (\ref{123}) on $\Omega (\chi)$, can be 
generalised further. In the generalised model also we 
take $\chi$ to have a finite upper bound 
$\chi \le \chi_{max} < \infty$ which was naturally 
the case in the previous model. Then, as in section 2,  
the factor $\chi_{max}$ can be absorbed into 
the Newton's constant $G_N$ and the range of $\chi$ can 
set to be $0 \le \chi \le 1$. The generalised model is 
now given by the function $\Omega (\chi) > 0$ satisfying 
the following more general constraints: 
\begin{eqnarray}
\Omega (0) & \le & \frac{1}{3}  \nonumber \\ 
\frac{d^n \Omega}{d \chi^n} & = & {\rm finite} 
\; \; \; \; \forall \; n \ge 1, \; \; \; \; 
0 \le \chi < 1  \nonumber \\ 
\Omega & \to & \infty \; \; \; \; {\rm at} 
\; \; \chi = 1 \; \; {\rm only} \nonumber \\ 
\lim_{\chi \to 1} \Omega & = & 
\Omega_1 (1 - \chi)^{- 2 \alpha} \; , \; \; \; \; 
\alpha > \frac{1}{4} \; , \label{123'} 
\end{eqnarray} 
where $\Omega_1 > 0$ is a constant. The function 
$\Omega (\chi)$ is arbitrary otherwise. 

Thus, in the generalised model, $\Omega (\chi)$ 
in the limit $\chi \to 0$ need not necessarily be 
a constant, but can be a function of $\chi$. For 
example, one may have $\Omega (\chi) = \Omega_{0'}  
\chi^{\beta}$ as $\chi \to 0$ where $\beta > 0$ and 
$\Omega_{0'} > 0$ are constants. Such functions do 
arise in the variable mass theories of \cite{mb} also. 
Note that the position of the pole, namely $\chi = 1$, 
is not a constraint but only a choice of normalisation 
of $\chi$. However, the order of the pole $\alpha$ can 
now be, for example, $\ge 1$. 

We have taken $\frac{d^n \Omega}{d \chi^n}$, 
$\forall n \ge 1$ for the sake of simplicity. This 
requirement can perhaps be relaxed if one uses only 
the appropraite necessary conditions to determine 
the absence of singularity instead of the sufficiency 
conditions given in the Appendix. However, we will not 
pursue it further in this paper. 

The evolution of the universe in the generalised model 
can be analysed along 
the same lines as in sections 4 - 6. It is clear 
that the only important limits are $\chi \to 0$ and, 
perhaps, $\chi \to 1$. Hence, we now analysis only these 
two limits in the generalised model. To be definite, we 
consider the following example: 
\begin{eqnarray}
\lim_{\chi \to 0} \Omega & = & \Omega_{0'} \chi^{\beta} 
\; , \; \; \; \; \beta > 0 \nonumber \\  
\lim_{\chi \to 1} \Omega & = & 
\Omega_1 (1 - \chi)^{- 2 \alpha} \; , \; \; \; \; 
\alpha \ge 1 \; , \label{omega01} 
\end{eqnarray} 
where $\Omega_{0'} > 0$ and $\Omega_1 > 0$ are 
constants. Note that the analysis in section 4 
applies directly with no modification in the limit 
$\chi \to 1$ if $\frac{1}{4} \le \alpha < 1$. Hence, 
we have taken $\alpha \ge 1$ in (\ref{omega01}). Using 
(\ref{omega}), the constraints (\ref{omega01}) can 
be expressed equivalently in terms of $\phi$ and 
$\psi (\phi)$ as follows: 
\begin{eqnarray}
\lim_{\psi \to - \infty} \psi & = & 
\frac{2}{\beta} {\rm ln} (\phi - \phi_0) \nonumber \\ 
\lim_{\psi \to 0} \psi & = & \psi_0 e^{- \phi} \; , 
\; \; \; \; {\rm if} \; \; \alpha = 1 \nonumber \\
& = & \psi_0 \; \phi^{- \frac{1}{\alpha - 1}} 
\; , \; \; \; \; {\rm if} \; \; \alpha > 1 \; , 
\label{psi01} 
\end{eqnarray} 
where $\phi_0$ and $\psi_0$ are constants. The range of 
$\psi = {\rm ln} \chi$ is given by 
$- \infty \le \psi \le 0$. Hence, the constant 
$\psi_0 < 0$ in (\ref{psi01}). Also, the limit 
$\chi \to 0$ corresponds to $\psi \to - \infty$ and, 
hence, to $\phi \to \phi_0$ whereas the limit 
$\chi \to 1$ corresponds to $\psi \to 0$ and, hence, 
to $\phi \to \infty$. 

\begin{center}
{\bf $\chi \to 0$} 
\end{center} 


In this limit, the analysis is similar to that given in 
section 4d. Under the conditions of section 4b, 
equation (\ref{achi}) can be solved in 
the limit $\chi \to 0$ relating $\chi$ and $e^A$: 
\begin{equation}\label{achi0'}
e^A = ({\rm constant}) \; \chi^{- \frac{1}{2}} \; . 
\end{equation}
Hence, $e^A \to \infty$ as $\chi \to 0$. Substituting 
(\ref{achi0'}) in equation (\ref{chid}) then yields 
the unique solution in the limit $\chi \to 0$: 
For $\beta \ne 1$, 
\begin{equation}\label{chi0'}
e^A = e^{A_0} \left( t_0 - {\rm sign} (m) t  
\right)^{- \frac{m}{2}} \; , \; \; \; \; \chi = 
\chi_0 \left( t_0 - {\rm sign} (m) t \right)^m \; , 
\end{equation}
where $t$ is measured in appropriate units, $A_0$, 
$\chi_0$, and $t_0 > t_i$ are constants and 
$m = \frac{2}{\beta - 1}$. For $\beta = 1$, 
\begin{equation}\label{chi0e'}
e^A = e^{A_0} e^{- \frac{c_e (t - t_0)}{2}} \; , 
\; \; \; \; \chi = \chi_0 e^{c_e (t - t_0)} \; , 
\end{equation}
where $c_e < 0$ is a constant. Note, as pointed out in 
\cite{lessner} and as discussed in section 3, that 
when $\chi$ and $e^A$ satisfies equation (\ref{achi0'}) 
the field equation (\ref{add}) is not guaranteed to 
follow from (\ref{ad}) - (\ref{rho}). Nevertheless, 
the above solution satisfies equation (\ref{add}) 
also, as checked explicitly. 

Let $\beta > 1$. Then, $m > 0$. Since $\chi \to 0$, it 
follows from (\ref{chi0'}) that $t \to t_0 > t_i$, 
implying that $\chi$ vanishes at a finite time $t_0$. In 
this limit, the scale factor $e^A \to \infty$ as follows 
from (\ref{chi0'}). The quantities in (\ref{qty}), for 
example $\frac{\dot{\chi}}{\chi}$, also diverge implying 
that the curvature invariants, including the Ricci 
scalar, diverge. Thus, for $\beta > 1$, there is 
a singularity at a finite time $t_0$. This implies, for 
example, that the variables mass theories of \cite{mb}
which exhibit a minimum of $e^A$ can nevertheless have 
singularities because $\beta$ can be $> 1$ in these 
theories. 

Let $\beta \le 1$. When $\beta < 1, \; m < 0$. Since 
$\chi \to 0$, it follows from (\ref{chi0'}) that 
$t \to \infty$. In this limit, the scale factor 
$e^A \to \infty$ as follows now from (\ref{chi0'}). 
The same result holds when $\beta = 1$ also, 
as follows from (\ref{chi0e'}). 

If $\rho \ne 0$ then the consistency of the above 
solutions requires that $\gamma \ge \frac{1}{3}$. This 
is analogous to the condition $\gamma > 0$ required 
in section 4d. 

It can now be seen, for $\beta \le 1$, that 
the quantities in (\ref{qty}) are all finite implying 
that all the curvature invariants are finite. Thus, 
there is no singularity in the above solutions when 
$\beta \le 1$. 

\begin{center}
{\bf $\chi \to 1$} 
\end{center} 


In this limit, the analysis is similar to that given in 
section 4a. For $\rho \ne 0$ and $\gamma \ne 1$, 
the solution is given uniquely by 
\begin{eqnarray}
e^A & = & e^{A_0} t^{\frac{2}{3 (1 + \gamma)}} 
\nonumber \\ 
\chi & = & 1 - \chi_0 e^{k T} \; , \; \; \; \; 
{\rm if} \; \; \alpha = 1 \nonumber \\
& = & 1 - \chi_0 (1 - k T)^{- \frac{1}{\alpha - 1}} 
\; , \; \; \; \; {\rm if} \; \; \alpha > 1 \; , 
\label{soln1'}
\end{eqnarray}
where $\chi_0 > 0$ and $k > 0$ are constants, $t$ is 
measured in appropriate units, and we have defined 
$T \equiv t^{- \frac{1 - \gamma}{1 + \gamma}}$ which 
$\to 0$ in the limit $t \to \infty$. The above solution 
then implies that $e^A \to \infty$ and $\chi \to 
(1 - \chi_0)$. The precise value of $\chi_0$ is model 
dependent. In particular, it also depends on $\Omega_1$ 
and can be made as small as necessary by choosing 
a sufficiently large $\Omega_1$. Equivalently, 
as $t \to \infty$, $\Omega$ can be made as large 
as necessary. 

This will imply, as in section 6, that in 
the present day universe in the generalised model, 
$\chi (today) \to 1 - \chi_0$, 
$\Omega (today) \to \Omega_1 \chi_0^{- 2 \alpha}$, and 
$\frac{1}{\Omega^3} \frac{d \Omega}{d \chi} (today) \to 
- \frac{2 \alpha}{\Omega_1^2} \chi_0^{4 \alpha - 1}$. 
Therefore, by choosing $\Omega_1$ sufficiently large 
one can ensure that the generalised model also satisfies 
the observational constraints imposed by solar system 
experiments, namely $\Omega (today) > 2000$ and 
$\frac{1}{\Omega^3} \frac{d \Omega}{d \chi} (today) 
< 0.0002$ (see \cite{will} pg. 117, 124-5, and 339). 

For $\rho = 0$, the solution is obtained easily in 
terms of $\phi$ and $\psi (\phi)$, and is given by 
\begin{eqnarray}
e^A & = & e^{A_0} t^{\frac{1}{3}} \nonumber \\ 
e^{\phi} & = & {\rm (constant)} \; 
t^{\frac{2}{\sqrt{3}}} \; , \label{rho0'} 
\end{eqnarray}
where $A_0$ is a constant. $\psi (\phi)$ and, hence, 
$\chi = e^{\psi}$ are then obtained from (\ref{psi01}). 
Note that, as in 
section 4a, these solutions also apply 
to the case when $\rho \ne 0$, but $\gamma = 1$. As 
$t \to \infty$ in the above solution, we have 
$e^A \to \infty$ and $\phi \to \infty$. Thus, as 
follows from (\ref{psi01}), $\psi \to 0$. Hence, 
$\chi \to 1$ and $\Omega \to \infty$. 

It can be seen that the quantities in (\ref{qty}) are 
all finite implying that all the curvature invariants 
are finite. Thus, there is no singularity in the above 
solutions. 

Now the evolution of the universe in the generalised 
model can be analysed as in sections 4 - 6. 
For the three toy universes of section 5, 
the evolution proceeds as described in section 5. 
For a realistic universe, such as our observed one, 
the evolution is complicated but proceeds as described 
in section 6. In particular, it follows that 
the evolution of the universe is singularity free in 
the generalised model when the function $\Omega (\chi)$ 
satisfies the constraints (\ref{omega01}). Although not 
proved here, the evolution is likely to be singularity 
free when $\Omega (\chi)$ satisfies the constraints 
(\ref{123'}) only. 

Before concluding this section, we would like to point 
out an interesting aspect of the generalised models, 
where $\Omega (\chi) \to 0$, equivalently $\omega \to 
- \frac{3}{2}$, in the limit $\chi \to 0$. When 
$\omega = - \frac{3}{2}$, the graviton-dilaton part 
of the action in (\ref{schi}) acquires a conformal 
symmetry under which 
\begin{equation}\label{confsym}
g_{\mu \nu} \to e^{\eta (x)} g_{\mu \nu} \; \; \; 
{\rm and} \; \; \; \chi \to e^{- \eta (x)} \chi \; , 
\end{equation} 
where $\eta (x)$ is an arbitrary function of space time 
coordinates \cite{wald}. In general, the matter action 
$S_M$ breaks this symmetry. However, it can be seen from 
the solutions presented in section 4 and 
the present one that precisely in the limit $\chi \to 0$ 
do the matter terms become negligible compared to 
the dilatonic terms. This suggests that the matter action 
$S_M$ may be neglected and that the conformal symmetry of 
the action (\ref{schi}) may become exact as $\chi \to 0$. 
A possible implication of the appearance of this 
conformal symmetry is mentioned in the conclusion.


\begin{center}
{\bf 8. Summary and Conclusion} 
\end {center} 


To summarise, we considered the evolution of a flat 
homogeneous isotropic universe in a graviton-dilaton 
model. The model is specified by the function 
$\Omega (\chi)$ which satisfies the constraints given 
in (\ref{123}), or more generally (\ref{123'}), but is 
arbitrary otherwise. We studied the evolution of 
universe in this model with the main purpose of 
determining whether it is singular or not. 

Assuming generic initial conditions where 
$\dot{A} (t_i)$ and $\dot{\chi} (t_i)$ are finite and 
non infinitesimal and $\Omega (t_i) < \infty$, we 
presented a general analysis which is valid for any 
matter and for any arbitrary function 
$\Omega (\chi)$, and 
which enables one to analyse the evolution and obtain 
its generic features even in the absence of explicit 
solutions. 

We illustrated our method by applying it to describe 
the evolution of three toy universes. We showed, in 
particular, that their evolution in the present model 
is singularity free. We then described the evolution 
of a universe, such as our observed one. We showed, 
in particular, that its evolution, although rich and 
complicated, is singularity free. This is a generic 
result valid for any $\Omega (\chi)$ satisfying only 
the constraints given in (\ref{123}) or (\ref{123'}). 

An important question to ask now, from our perspective, 
is whether a function $\Omega (\chi)$ as required in 
the present model can arise from a fundamental theory. 
We now discuss critically a few such possibilities in 
string theory. 

In the string theoretic context, Einstein metric 
formulation given in (\ref{sein}) is more natural where 
$\phi$ is the dilaton and $\chi (\phi)$ the arbitrary 
function. The strong coupling limit in string theory, 
relevant for singularities, corresponds to diverging 
effective Newton's constant ($= \frac{1}{8 \pi \chi}$) 
and, thus, to the limit $\chi (\phi) \to 0$. In this 
limit ${\rm ln} \chi \to \frac{\phi}{\sqrt{\Omega_0}}$ 
in our model (see equation (\ref{123}) and (\ref{sein})). 

(1) Note that, in string theory, ${\rm ln} (\chi (\phi))$ 
can be thought of as Kahler potential for $\phi$, which 
is expected to be modified at strong coupling by 
non perturbative effects \cite{quevedo}. But, such 
modifications are usually exponential in nature and, 
hence, are unlikely to lead to a $\chi (\phi)$ as 
required here. 

(2) However, at strong coupling, $\phi$ here may not be 
the stringy dilaton directly, but instead a combination of 
the stringy dilaton and other compactification dependent 
moduli. In that case, after writing the relevant effective 
action as given in (\ref{sein}), the Kahler potential 
${\rm ln} \chi$ for this `effective dilaton' $\phi$ 
may well turn out to be of the required form. 

(3) Another, perhaps more promising, 
possibility is the following which 
arises when $\Omega_0 = \frac{1}{3}$ (see equation 
(\ref{123}) and (\ref{sein})). This case corresponds to 
a five dimensional space time compactified to four 
dimensions on a circle \cite{gm}. In recent developments 
in string theory at strong coupling, analogous phenomena 
relating a $d$-dimensional theory and 
a $(d + 1)$-dimensional theory on a circle are found to 
occur: Using S-duality symmetries, Witten has discovered 
that the ten dimensional string at strong coupling is 
an eleven dimensional (M-)theory compactified on a circle 
\cite{w,rev}. A similar phenomenon, 
that the three dimensional string at strong coupling 
is a four dimensional theory compactified on a circle, 
is at the heart of Witten's novel proposal for solving 
the cosmological constant problem \cite{cc}; for its 
possible stringy realisation see \cite{v}. A similar 
phenomenon in four/five dimensions, if exists, may 
possibly lead to a $\chi (\phi)$ as required here. 

(4) As discussed in section 7, 
the graviton-dilaton action in (\ref{schi}) acquires 
a conformal symmetry in the limit $\chi \to 0$ 
where $\Omega (\chi) \to 0$, equivalently 
$\omega (\chi) \to - \frac{3}{2}$. Note that this is 
also the limit where singularities can potentially arise, 
but are avoided in the present model. The apperance 
of this conformal symmetry is, perhaps, a hint of 
a connection between the present model and string 
theory, known to lead generically to conformal field 
theories, but which must now be in the strong coupling 
limit since $\chi \to 0$. 

Admittedly, at present, these are plausibility arguments 
only. Nevertheless, given the elegant way the present 
model resolves the big bang singularity it is worthwhile 
to derive it, perhaps along the above lines, from 
a fundamental theory such as string theory. 

On the other hand, however, such a model, even if not 
derivable from string theory, may stand on its own as 
an interesting model for singularity free cosmology. One 
may then study its implications in other cosmological 
and astrophysical contexts. There are many issues that 
can be studied further. We mention some of them here. 

(1) One can study the evolution of a more complicated 
universe, {\em e.g.} an inhomgeneous anisotropic 
universe, and determine whether it is singularity free 
or not. 

(2) In section 6 we found that the evolution of 
a realistic universe can be oscillatory. Such early 
universe oscillations have significant implications and 
may also lead to observable predictions. It is therefore 
important to study this issue in more detail. 

(3) In the astrophysical context, one may study 
the evolution and/or collapse of stars. In fact, in 
\cite{k1} where the present model was originally derived 
in a different context, we found that our model suggests 
a novel scenario for stellar collapse in which a black 
hole is unlikely to form. Such a phenomenon, if 
established rigorously, is likely to have far 
reaching consequences. 

(4) The present model is characterised by one arbitrary 
function $\Omega (\chi)$, required to satisfy 
the constraints given in (\ref{123}) or (\ref{123'}). 
These constraints, however, do not fix $\Omega (\chi)$ 
uniquely. The consequent arbitrariness in the model 
greatly diminishes its predictive power. It is 
therefore desirable to find the criteria which 
can fix $\Omega (\chi)$ uniquely. 

In this regard, note that the present model naturally 
incorporates, as described in \cite{k2}, an ingredient 
crucial for the success of hyperextended inflation 
\cite{stein}. Therefore, with a view to fix 
$\Omega (\chi)$ uniquely, one may require the present 
model to lead to a successful inflation also. However, 
our preliminary calculations \cite{ks} indicate that this 
requirement is not strong enough to fix $\Omega (\chi)$ 
uniquely. Hence, more criteria are needed. 

Considering the simplicity of the present model and 
the novel consequences that follow from it generically, 
we believe that its further study is fruitful and is 
likely to lead to interesting phenomena. 

\vspace{2ex}

{\bf Acknowledgement:} 
We thank T. R. Seshadri for discussions. 


\begin{center}
{\bf APPENDIX}

\vspace{2ex}

{\bf Finiteness of Curvature Invariants} 
\end {center} 


In this Appendix, we derive a sufficient condition 
for {\em all} curvature invariants to be finite. 
Curvature invariants are constructed using metric tensor, 
Riemann tensor, and covariant derivatives. When the metric 
is diagonal, every term in any curvature invariant can be 
grouped into three types of factors: 
(A) $\sqrt{g^{\mu \mu} g^{\nu \nu} g^{\lambda \lambda} 
g^{\tau \tau}} \; R_{\mu \nu \lambda \tau}$, 
(B) $\sqrt{g^{\mu \mu } g^{\nu \nu } g_{\lambda \lambda}} 
\; \Gamma_{\mu \nu}^\lambda$, and 
(C) $\sqrt{g^{\mu \mu}} \; \partial_\mu$ acting 
multiply on (A) and (B) type factors (no summation over 
repeated indices). 

Using the metric given by (\ref{ds}) and evaluating 
the above factors explicitly, one obtains that (A) and 
(B) type factors are functions of $\ddot{A}, \dot{A}$, 
and $e^{- A}$. The action of (C) produces extra time 
derivatives. Thus, it follows that the curvature 
invariants are functions of $e^{- A}$ and 
$\frac{d^n A}{d t^n}, \; n \ge 1$. 

By repeated use of equations (\ref{ad})-(\ref{rho}), 
$\frac{d^n A}{d t^n}$ for any $n \ge 1$ and, hence, 
any curvature invariant, can be expressed in terms of 
the following quantities: 
\begin{equation}\label{qty}
e^{- A}; \; \; \; 
\frac{\rho}{\chi}, \;  
\frac{\rho}{\chi \Omega}; \; \; \; 
\frac{\dot{\chi}}{\chi}, \; 
\frac{\Omega \dot{\chi}^2}{\chi^2}; \; \; \; 
{\rm and} \; \; \; 
\frac{\chi^n}{\Omega} \; \frac{d^n \Omega}{d \chi^n} \; 
\left( \frac{\dot{\chi}}{\chi} \right)^n, \; \; 
\forall n \ge 1 \; . 
\end{equation}
The resultant expressions are finite if the above 
quantities are finite. The algebra is straightforward 
and, hence, we omit the details here. 

Hence, {\em a sufficient condition for all curvature 
invariants to be finite is that the quantities in 
(\ref{qty}) be all finite}.

\end{document}